\documentclass[10pt]{iopart}
\usepackage{iopams}
\usepackage[utf8]{inputenc}
\usepackage[english]{babel}
\usepackage[pdftex]{graphicx}
\usepackage{stfloats}
\usepackage{float}
\usepackage{enumitem}
\usepackage{url}
\usepackage[nolist]{acronym}
\usepackage[binary-units=true, separate-uncertainty, separate-uncertainty-units = single]{siunitx}
\usepackage{booktabs}
\usepackage{subfig}
\usepackage{lineno}

\usepackage[hidelinks]{hyperref}
\usepackage[english, nameinlink, capitalise, noabbrev]{cleveref}
\usepackage{acronym}
\usepackage{todonotes}

\usepackage{amsfonts}
\usepackage{rotating}
\usepackage{pdflscape}
\usepackage{microtype}
\usepackage{multirow}

\usepackage{scalerel}
\usepackage{tikz}
\usetikzlibrary{svg.path}
\definecolor{orcidlogocol}{HTML}{A6CE39}
\tikzset{
  orcidlogo/.pic={
    \fill[orcidlogocol] svg{M256,128c0,70.7-57.3,128-128,128C57.3,256,0,198.7,0,128C0,57.3,57.3,0,128,0C198.7,0,256,57.3,256,128z};
    \fill[white] svg{M86.3,186.2H70.9V79.1h15.4v48.4V186.2z}
                 svg{M108.9,79.1h41.6c39.6,0,57,28.3,57,53.6c0,27.5-21.5,53.6-56.8,53.6h-41.8V79.1z M124.3,172.4h24.5c34.9,0,42.9-26.5,42.9-39.7c0-21.5-13.7-39.7-43.7-39.7h-23.7V172.4z}
                 svg{M88.7,56.8c0,5.5-4.5,10.1-10.1,10.1c-5.6,0-10.1-4.6-10.1-10.1c0-5.6,4.5-10.1,10.1-10.1C84.2,46.7,88.7,51.3,88.7,56.8z};
  }
}
\newcommand{\orcidicon}[1]{\href{https://orcid.org/#1}{\mbox{\scalerel*{
\begin{tikzpicture}[yscale=-1,transform shape]
\pic{orcidlogo};
\end{tikzpicture}
}{|}}}}

\usepackage[backend=bibtex, bibencoding=utf-8, style=ieee, sorting=none, doi=true, url=false, isbn=false, style=numeric-comp]{biblatex}

\usepackage{geometry}
\usepackage[babel]{csquotes}
\uspunctuation
\begin{acronym}
\acro{GATE}{GEANT4 Application for Tomographic Emission}
\acro{GAN}{generative adversarial network}
\acro{PET}{positron emission tomography}
\acro{cGAN}{conditional generative adversarial network}
\acro{BGO}{Bismuth Germanate}
\acro{SiPM}{silicon photomultiplier}
\acro{DOI}{depth-of-interaction}
\acro{SWD}{sliced Wasserstein distance}
\acro{SWS}{sliced Wasserstein similarity}
\acro{COG}{center-of-gravity}
\acro{SSIM}{structural similarity index measure}
\acro{SPAD}{single photon avalanche diode}
\acro{LSO}{Lutetium Oxyorthosilicate}
\acro{MAE}{mean-absolute-error}
\end{acronym}

\begin{document}
%\linenumbers
\title[Physics-Informed Generative Modeling of Optical Photon Transport]%
      {OptiGAN for Crystal Arrays: Physics-Informed Generative Modeling
       of Optical Photon Transport in PET Detector Arrays%
       \footnote{Accepted for publication in \textit{Physics in Medicine
       \& Biology}, DOI: \href{https://doi.org/10.1088/1361-6560/ae95d7}%
       {10.1088/1361-6560/ae95d7}. Please cite the published version.}}
\author{Stephan Naunheim${^1}$\orcidicon{0000-0003-0306-7641}, Brandon Pardi${^2}$\orcidicon{0000-0001-6483-9858}, Guneet Mummaneni${^2}$\orcidicon{0009-0006-5745-7249}, Carlotta Trigila${^{1}}$, Emilie Roncali$^{{1}}$\orcidicon{0000-0002-2439-1064}}
\address{$^1$Department of Biomedical Engineering, University of California, Davis, Davis, CA, United States of America}
\address{$^2$Department of Computer Science, University of California, Davis, Davis, CA, United States of America}
\ead{snaunheim@ucdavis.edu}

\begin{abstract}
\newline
\textit{Objective.}
Monte Carlo simulations of optical photon transport are computationally prohibitive for large-scale optical systems including detector arrays and PET systems, restricting their practical use to single-crystal studies. This work presents an enhanced conditional generative adversarial network capable of replacing optical simulations at the crystal array level, extending our previous single-crystal approach to a \numproduct{3x3} BGO detector array.\newline
\textit{Approach.}
We introduce Fourier feature encoding and a learnable latent mapping network as the modifications enabling stable training on the array geometry, together with a physics-informed loss term enforcing the unit-sphere state space $S^2$ of the generated propagation directions as a soft constraint. Training data requirements are reduced eight-fold by exploiting the array's symmetry. The model is evaluated in three studies: a full array evaluation, a high-resolution study probing generalization, and a pencil beam irradiation study assessing practical applicability. Performance is benchmarked against GATE10/Geant4 ground truth, using the fluctuations between independent Monte Carlo runs.\newline
\textit{Main results.}
The enhanced optiGAN achieves similarity values within $3\sigma$-agreement of the Monte Carlo baseline across all evaluation conditions. An ablation and attribution analysis shows that the physics-informed loss term reduces low-SSIM bin fractions by a factor of \num{3.6} on the outer crystals, with a localized trade-off at the central crystal, yielding a net \qty{48}{\percent} reduction over the full array. The model transitions from electron-emission training data to realistic $\gamma$-photon interactions, producing flood maps that reproduce experimental patterns including photopeak clusters and inter-crystal scatter lines.\newline
\textit{Significance.}
This proof-of-concept demonstrates that a physics-informed generative model can simulate optical photon transport in segmented scintillator arrays at a training and inference cost accessible on a single workstation GPU. This provides a foundation for future models capable of generalizing across diverse array configurations.\\\par 

\end{abstract}

\maketitle

\acresetall

\section{Introduction}
\Ac{PET} is a medical imaging technique \cite{phelps_application_1975, bailey_positron_2005, cherry_physics_2012} that aims to provide functional information on physiological processes within a subject. Imaging relies on the registration of coincident annihilation quanta using dedicated radiation detectors, which often consist of scintillation crystals coupled to readout sensors \cite{berg_innovations_2018}. While \ac{PET} research is multifaceted, one major branch focuses on advances in detection instrumentation to improve the information available for image reconstruction. Typically, Monte Carlo simulations \cite{sarrut_advanced_2021, sarrut_gate_2025, krah_gate_2025} are used to evaluate the feasibility of, for example, novel detector concepts or data-processing algorithms. Although these simulations represent the gold standard, the time required to run them is often long due to stepwise particle tracking and, therefore, prohibitive for large-scale detector system setups.\newline
Furthermore, one is often not interested in every detail of individual particle tracks (e.g., reflections, transmission, binary scattering flag), but rather in the signals produced by the sensor that can be used in experiments. One approach to reduce simulation time is to use generative models \cite{de_oliveira_learning_2017, paganini_accelerating_2018, paganini_calogan_2018, sarrut_generative_2019, fanelli_machine_2020, fanelli_deeprich_2020, sarrut_modeling_2021, trigila_generative_2023, naunheim_novel_2026} that directly produce information at the readout sensor level and do not rely on tracking particles through the detection medium. This new approach potentially offers new insights into detector modeling by allowing the user to generate only signals and data that satisfy specified conditions, which might be challenging to achieve in classical Monte Carlo simulations constrained by physical laws. However generative models must be trained to learn the relevant statistical distributions, which requires time and the availability of training data that provide de facto ground truth. For this reason, the primary application of generative AI modeling  will likely be at the detector and system levels rather than at the level of single crystals. \newline 
In previous research \cite{trigila_generative_2023, trigila_towards_2025, mummaneni_optigan_2025}, our group demonstrated the feasibility of building a conditional \ac{GAN} \cite{goodfellow_generative_2014, mirza_conditional_2014, arjovsky_wasserstein_2017}, called optiGAN, that can replace optical simulations at the single crystal level. Furthermore, we integrated the model into the open-source GATE10 ecosystem, enabling interested users to test and work with it.\newline
In this new proof-of-concept study, we present a novel model in the optiGAN family designed to operate on crystal arrays rather than single crystals. Although the current implementation, including training overhead, does not outperform conventional optical simulations in total computation time yet, this work provides essential insights into array-level generative modeling. The \ac{GAN} can be used to generate data representing the detection of optical photons at the attached sensor surfaces based on the conditioned location of the $\gamma$-photon interaction in the crystals $\mathbf{y} = (X_{em},Y_{em},Z_{em})$. The generation includes the spatial coordinates ($\mathbf{R}_i = (x_i, y_i)^T$) of an impinging optical photon $i$, its energy ($E_i$), and the time of detection ($T_i$). In addition, the model generates the normalized directional momentum components ($\hat{\mathbf{P}}_i = (p_{x,i}, p_{y,i}, p_{z,i})^T$), which are not directly relevant in experimental settings but provide a means to introduce physics-informed knowledge into the model.
Compared with our previously published OptiGAN model, we introduce significant architectural changes to account for the increased complexity of modeling crystal arrays rather than single crystals while maintaining computational accessibility for typical GATE users. The network modifications are justified through a sequential ablation study. We augment the latent vector with Fourier components \cite{rahimi_random_2007, tancik_fourier_2020} with the aim of helping the model learn non-smooth, step-wise distributions, which are expected to be seen at the boundaries between neighboring crystals. While this approach greatly enriches the latent representation, it also drastically increases the dimensionality of the latent space, which is a non-desired side effect for our use case. Therefore, we employ a small learnable mapping network \cite{karras_style-based_nodate} as a preprocessor for the generator input, selecting and weighting the important components of the augmented latent vector to compress the input to the generator's original dimensionality. Following this strategy, we can build more complex models without a significant increase in computational cost relative to our previously published single-crystal model. The last significant modification addresses the adversarial loss function by adding a physics-informed loss term that softly enforces the unit-norm property of the generated normalized momentum vectors. We kept the parameter count similar to the original architecture to ensure that users without access to high-performance computing resources can still use it effectively. \newline
We apply the optiGAN model to simulated data of a \ac{BGO} array consisting of \numproduct{3 x 3} crystals, each with a volume of \qtyproduct{3 x 3 x 10}{\milli \metre}. The model is trained on data from only one eighth of the array's quadratic base area (fundamental domain). Still, we test its capabilities with distinct studies using the full-crystal array, a high-resolution study, and a practical use case with pencil-beam irradiation.

\section{Materials}

\subsection{Physics-informed OptiGAN with Latent Fourier Augmentation}
\label{subsec:pi-intro}

\begin{figure}
    \centering
    \includegraphics[width=0.9\linewidth]{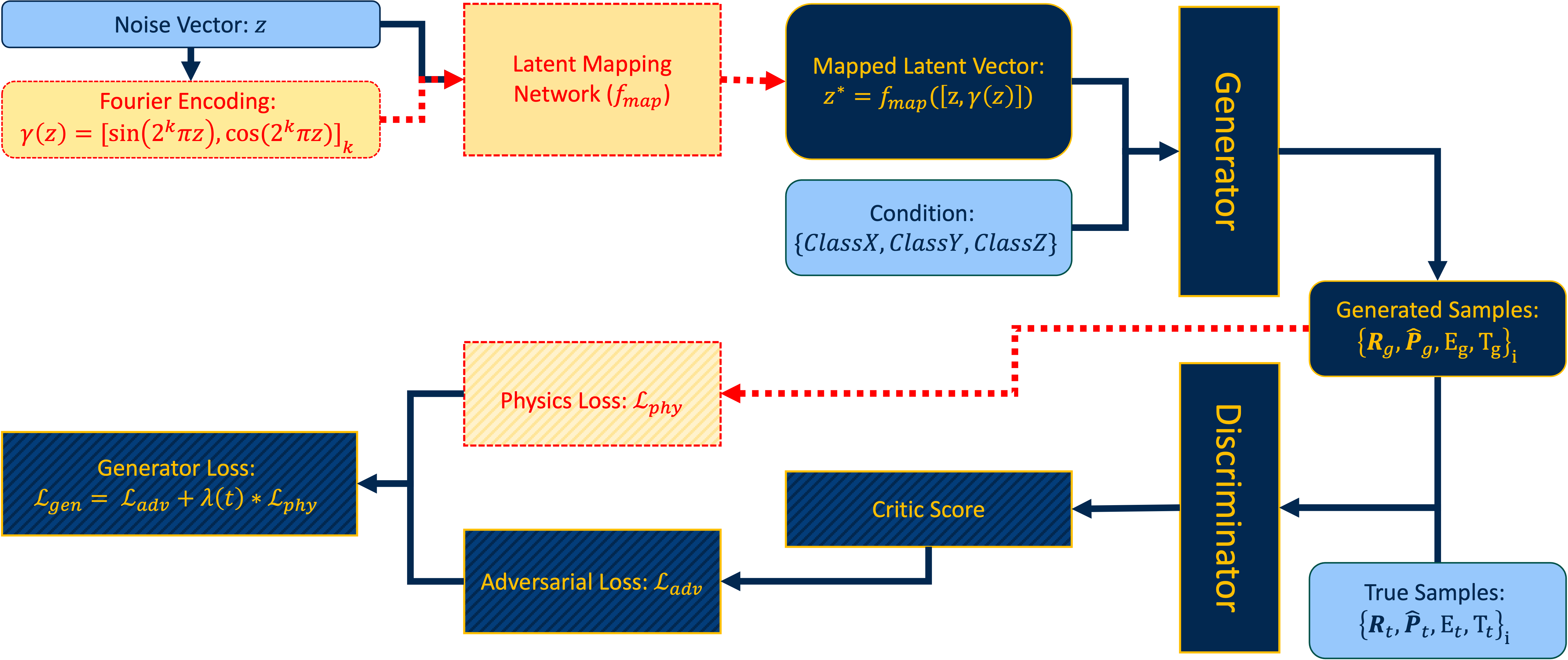}
    \caption{Schematic architecture of modified optiGAN model. Rectangles in yellow represent newly introduced modifications for properly modeling the crystal array. Input data is marked as light blue rectangles. The generator and discriminator design have not been changed.}
    \label{fig:NetworkArch}
\end{figure}

We extend the classical OptiGAN architecture \cite{trigila_generative_2023, trigila_towards_2025, mummaneni_optigan_2025}, which follows a Wasserstein-\ac{GAN} \cite{arjovsky_wasserstein_2017} design with gradient penalty \cite{gulrajani_improved_2017}, with two key innovations to improve sample quality and enforce physical constraints (see \cref{fig:NetworkArch}) for properly modeling the crystal array. The innovations include changes in the latent space and loss function, such that the core network architecture of the generator and discriminator, which worked for single crystals, remains unchanged. Each modification introduces a new set of hyperparameters. Given the proof-of-concept nature of this study, we did not perform an extensive hyperparameter optimization. Instead, a small set of commonly used hyperparameter values was evaluated to validate the core idea. During the training procedure, we monitored the \ac{SWS} (see \cref{subsec:dist_sim}) performance on unseen test data. The best-performing checkpoint is defined as the epoch with the highest condition-wise average of the \ac{SWS}. A systematic ablation of the individual contributions of these modifications is presented in \cref{subsec:ablation_results} and forms the basis for the selection of the final model configuration.

\paragraph{Fourier Feature Encoding \& Latent Mapping Network} 
To address the spectral bias of neural networks, which struggle to learn high-frequency functions, we augment the input latent vector $\mathbf{z} \in \mathbb{R}^d$ with Fourier feature encoding \cite{tancik_fourier_2020},
\begin{equation}
\gamma(\mathbf{z}) = \left[\sin(2^k \pi \mathbf{z}), \cos(2^k \pi \mathbf{z})\right]_{k=0}^{n_{\text{freq}}-1} \in \mathbb{R}^{2nd},
\end{equation}
where $n_{\text{freq}}$ controls the number of frequency components. This transformation provides the network with explicit multi-scale representations, enabling it to capture fine-grained details and sharp features that are difficult to learn from raw noise alone.\newline
In this work, we follow our prior publications and use $d=7$ original latent features. The number of Fourier frequencies is set to $n_{\text{freq}} = 4$, based on the results of a preliminary limited hyperparameter search.\newline
Rather than directly passing the high-dimensional noise vector to the generator, which heavily increases computational load, we employ a learnable latent mapping network \cite{karras_style-based_nodate} that adaptively combines the original latent vector and its Fourier encoding,
\begin{equation}
\mathbf{z}^* = f_{\text{map}}([\mathbf{z}, \gamma(\mathbf{z})]; \theta_m),
\end{equation}
where $f_{\text{map}}$ is a two-layer MLP with LeakyReLU activation that maps the concatenated $d + 2nd$ dimensional input back to $d$ dimensions. This network learns to weight and combine frequency components optimally for the generation task, reducing dimensionality while preserving important spectral information.\newline
%An initial small ablation study indicates that a model modified with the Fourier feature encoding and latent mapping network shows improved training stability compared to the optiGAN design used for single crystals.

\paragraph{Physics-Informed Loss}

Depending on the emission location of the optical photons, the attached sensors capture different distributions of photon momenta due to geometric constraints, e.g., reflections. Those distributions carry physical meaning, since they show allowed and forbidden regions in the momentum space. In GATE, the momentum phase space of each optical photon is factorized as the scalar kinetic energy E, and the normalized momentum components (dX, dY, dZ), which are confined to the unit sphere $S^2$. Both factors carry physical information, as E fixes the momentum magnitude via $\left|\boldsymbol{p}\right|=E/c$ for a massless particle, while the normalized components encode the geometric propagation direction $\hat{\boldsymbol{P}}=\left(dX,dY,dZ\right)$ shaped by reflection and refraction within the crystal array. A generative model of this phase space must therefore preserve the $S^2$ constraint on its direction output, which is why we use the unit-norm loss enforcing this as a soft penalty \cite{karniadakis_physicsinformed_2021, jin_physicsinformed_2022} during training. For photon momenta $\hat{\mathbf{P}}_i$ in a batch $\mathcal{B}$, the physics loss $\mathcal{L}_{\text{phy}}$ is defined as
\begin{equation}
\mathcal{L}_{\text{phy}} = \mathbb{E}_{i \sim \mathcal{B}}\left[ \left( \|\hat{\mathbf{P}}_i\|^2 - 1 \right)^2 \right],
\label{eq:physics_loss}
\end{equation}
which penalizes deviations of each generated direction vector from unit norm. This loss is combined with the adversarial WGAN-GP objective $\mathcal{L}_{\text{adv}}$ using an annealing schedule $\lambda(t)$,
\begin{equation}
\mathcal{L}_{\text{gen}} = \mathcal{L}_{\text{adv}} + \lambda(t) \cdot \mathcal{L}_{\text{phy}},
\label{eq:gen_loss}
\end{equation}
that linearly increases from 0 to the target weight over a warmup period, allowing the generator to first learn the overall data distribution before enforcing physical constraints. We chose a warmup period of $5000$ epochs, with a target weight of $\lambda(t \geq 5000) = 0.01$. The complete generator architecture processes the mapped latent representation $\mathbf{z}^*$, concatenated with the condition labels $\mathbf{y}$, through a four-layer MLP with ReLU activations to produce the final output. The discriminator follows a similar architecture with dropout regularization ($p=0.1$) to prevent overfitting.\newline
Within the ablation study, we denote the single-crystal baseline architecture as optiGAN, the model with Fourier feature encoding and mapping network as optiGAN+Fourier, and the full model with unit-norm loss as optiGAN+Fourier+PI.

\subsection{PET Detector \& General Simulation Settings}

\begin{figure}
    \centering
    \includegraphics[width=0.4\linewidth]{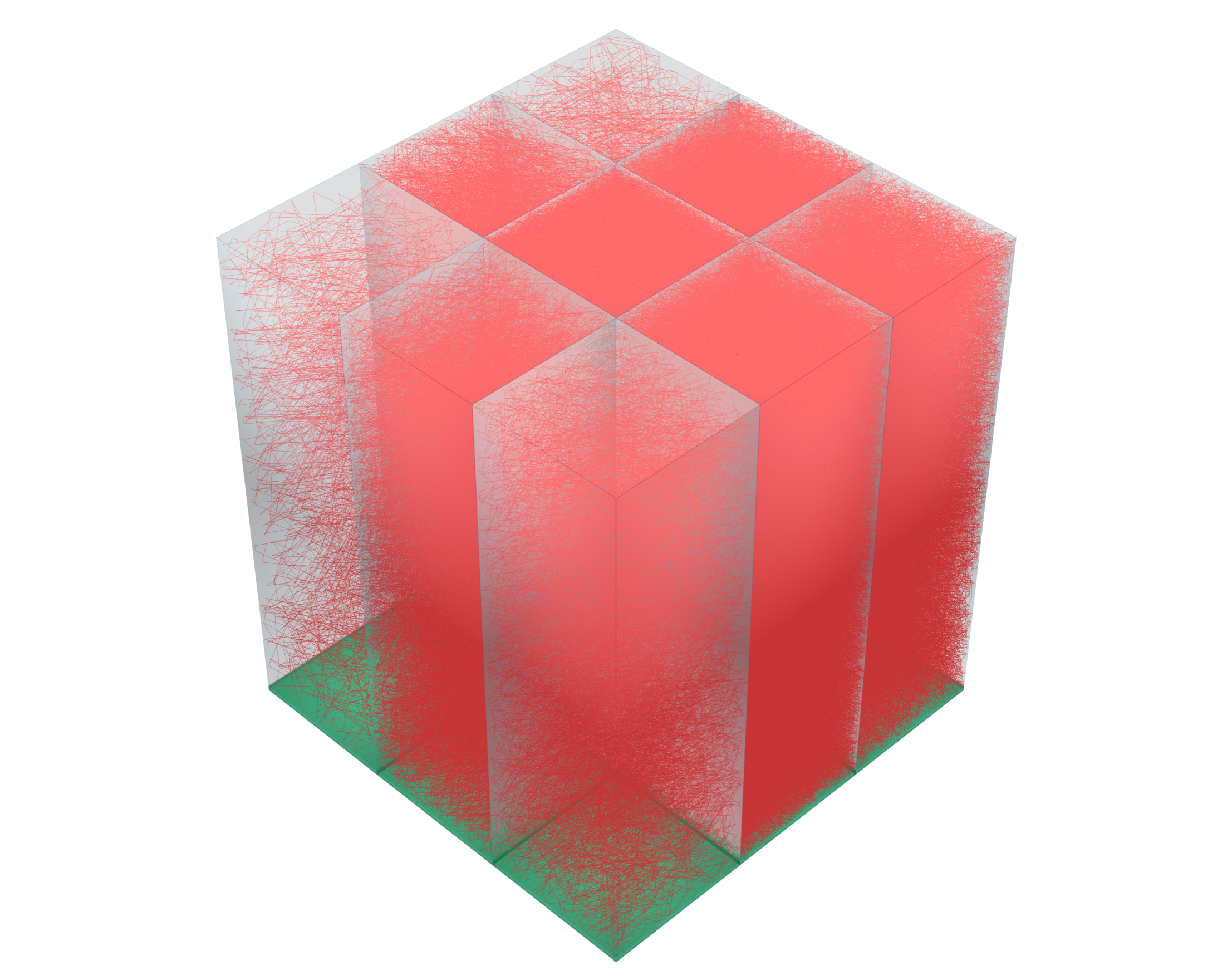}
    \caption{Simulated crystal array consisting of \numproduct{3 x3} \ac{BGO} crystals.}
    \label{fig:Array}
\end{figure}

The simulated \ac{PET} detector (see \cref{fig:Array}) is comprised of \numproduct{3x3} \ac{BGO} crystals. Each crystal has a base area of \qtyproduct{3 x 3}{\milli \metre} and a height of \qty{10}{\milli \metre}. Each crystal is coupled to an \ac{SiPM}, modeled as a \qty{200}{\micro \metre} thick SiO$_2$ layer. The optical coupling between the crystal and \ac{SiPM} is given as an Epoxy layer with a thickness of \qty{200}{\micro \metre}. All crystals have polished surfaces and are optically separated from one another using Teflon (diffuse reflection). Optical surfaces are modeled using the Davis LUT model \cite{roncali_simulation_2013, roncali_cerenkov_2019, trigila_integration_2021}.\newline
Simulations are conducted with GATE10 (version 10.0.2) \cite{sarrut_gate_2025, krah_gate_2025} using the \texttt{G4EmStandardPhysics\_option4} physics list. The production cuts within the crystals are set to \qty{10}{\micro \metre}. Phasespace actors attached to the crystals and sensor volumes are used to cluster optical photon data from the same initial interaction event.

\subsection{Fundamental-Domain Training Dataset}
\label{subsec:fundamentalDomain}
The crystal array's square geometry induces a $D_4$ point group symmetry. This symmetry manifests as invariance under $90^{\circ}$ rotations and mirror reflections, which translates to the absence of preferential light propagation directions. By exploiting this symmetry, it is possible to limit the training data creation to the so-called fundamental domain $\mathcal{F}$ of the array's base area,
\begin{equation}
    \mathcal{F} = \{ (x,y): 0 \leq y \leq x \leq L/2 \},
\end{equation}
with $L$ denoting the side length of the array. Any point outside this region can be mapped to it through an appropriate symmetry operation $g \in D_4$ to a point in the fundamental domain. Since $\mathcal{F}$ covers only about an eighth of the full array area, this technique allows an eight-fold reduction in computational time and memory requirement.\newline
The isotropic emission of mono-energetic electrons ($E_{e^-}=\qty{420}{\kilo \eV}$) at specific locations within the crystals is simulated to create training and testing data. The choice of \qty{420}{\kilo \eV} reflects the kinetic energy of a photoelectron produced by photoelectric absorption of a \qty{511}{\kilo \eV} annihilation quantum on the K-shell of bismuth, which is the dominant photoelectric channel in \ac{BGO} at \ac{PET}-relevant energies, and is consistent with our previous single-crystal optiGAN studies \cite{trigila_generative_2023, mummaneni_optigan_2025, trigila_towards_2025}. The training data themselves contain only the optical photons recorded at the sensor surfaces, such that the primary electron energy is not included in the conditioning vector. For other scintillators such as \ac{LSO}, the same methodology applies with an adapted primary electron energy.\newline
This study is the first translation of the optiGAN methodology from single-crystal to arrays, therefore we opted to keep parameters consistent with the prior single-crystal work \cite{trigila_towards_2025, mummaneni_optigan_2025}  as much as possible. The emission locations are thus distributed on a regular grid over the fundamental domain, repeated at three $Z$-layers at heights of \qty{1}{\milli \metre}, \qty{5}{\milli \metre}, and \qty{9}{\milli \metre} from the sensor surface. On the single \qtyproduct{3x3x10}{\milli \metre} crystal, a step size of \qty{0.5}{\milli \metre} was found sufficient for target-distribution convergence, and further refinement did not measurably improve fidelity. Translating this grid to the \numproduct{3x3} array is not straightforward, since a step size of \qty{0.5}{\milli \metre} would place emission locations exactly at inter-crystal boundaries, where the simulation is ill-defined. Refining below \qty{0.5}{\milli \metre} is not empirically motivated either, since the array already covers a nine-fold larger base area than the single-crystal reference at comparable density. We therefore chose \qty{0.6}{\milli \metre} as the smallest boundary-avoiding step size close to the value validated on the single-crystal geometry. In total, this yields 108 training conditions in $\mathcal{F}$, with approximately $3 \times 10^4$ optical photons recorded per emission location.

\subsection{Hardware \& Computational Performance}
All Monte Carlo simulations were performed on an AMD Ryzen Threadripper 7960X CPU. \Ac{GAN} training and inference were performed on a NVIDIA RTX 4080 SUPER GPU. The per-condition wall-clock time for the Monte Carlo simulation of 100 primary \qty{420}{\kilo \eV}-electrons was \qty{312.1 \pm 10.3}{\second} ($n = 5$ conditions randomly sampled from the training grid). The complete training dataset (108 conditions) thus required approximately \qty{9.4}{\hour} of single-core wall-clock time. The \Ac{GAN} training proceeded at approximately \qty{13}{\second} per epoch, corresponding to about \qty{70}{\hour} for the modified optiGAN model. The \Ac{GAN} inference for an output volume equivalent to one Monte Carlo simulation required \qty{3.73 \pm 0.01}{\milli \second} per condition on the same GPU ($n = 20$ runs, batch size 500,000), corresponding to a throughput of approximately $22.8 \times 10^6$ samples per second.

\section{Methodology}

\subsection{Distributional Similarity}
\label{subsec:dist_sim}

To quantitatively evaluate the quality of generated samples, we employ the \ac{SWD}, which efficiently approximates the high-dimensional Wasserstein distance via random one-dimensional projections. This metric captures differences in the full multivariate distribution structure while remaining computationally tractable, and provides stable gradients even for non-overlapping distributions.

To obtain a normalized similarity metric in $[0,1]$, we compute a baseline distance between the real data distribution and a Dirac delta distribution centered at zero ($D_{\text{baseline}} = \text{SWD}(\mathbf{p}_{\text{real}}, \delta_0)$). The normalized \ac{SWS} is then defined as
\begin{equation}
\text{SWS} = 1 - \frac{\text{SWD}(\mathbf{p}_{\text{real}}, \mathbf{q}_{\text{gen}})}{D_{\text{baseline}}},
\end{equation}
where a value of 1 indicates a perfect distributional match. This normalization makes the metric interpretable and invariant to feature scale.\newline
In our implementation, we use $K=60$ random projections and subsample both distributions to 50,000 points. Rather than evaluating performance globally, we compute the \ac{SWS} for each condition separately to assess the generator's accuracy across different spatial regions within the crystal array, enabling identification of regions where the generator performs well or requires improvement.

\subsection{Visual Similarity}
\label{subsec:visual_sim}

From an instrumentalist perspective, the detailed distributional analysis presented in \cref{subsec:dist_sim} addresses quantities that are often difficult to access experimentally. Momentum distributions, for instance, are not measured in typical \ac{PET} systems, and the optical photon signals themselves undergo multi-step transformations, including applying the \ac{SiPM} response, which inhibits photon detection efficiency, dark counts, and electronic noise before reaching the data acquisition system. Consequently, for practical applications, it is more meaningful to evaluate the agreement using metrics that reflect standard experimental characterization procedures employed in instrumentation research. Therefore, we compare GAN-generated and Monte Carlo flood maps using the \ac{SSIM} \cite{wang_image_2004}. Unlike pixel-wise metrics, the \ac{SSIM} accounts for the perceptual quality of image structures and is particularly sensitive to the preservation of shapes, edges, and spatial patterns.\newline
In practice, flood maps are the standard tool for assessing crystal identification and separation quality in pixelated scintillator arrays, as they distinguish between photopeak and inter-crystal scatter events \cite{zhang_design_2021, refaey_optimizing_2024, cong_algorithms_2025}. They are generated by binning the \ac{COG} positions computed for each recorded $\gamma$-photon. For the $k$th $\gamma$-photon producing $N_k$ detected optical photons, the \ac{COG} is calculated as the energy-weighted mean detection position,
\begin{equation}
\mathbf{R}_{\mathrm{COG},k} = \frac{\sum_{i=1}^{N_k} E_i \, \mathbf{R}_i}{\sum_{i=1}^{N_k} E_i},
\end{equation}
where $\mathbf{R}_i$ and $E_i$ denote the detection position and energy of the $i$th optical photon, respectively. Those maps are inherently sparse, since \ac{COG} positions cluster typically around crystal centers and along inter-crystal scatter lines while the intervening regions remain empty. These empty regions are just as physically informative as their non-empty counterparts. A metric restricted to non-empty bins would not penalize spurious counts such as ghost peaks or excessive cluster spread. We therefore evaluate the \ac{SSIM} over the full flood map area, retaining the empty regions as part of the evaluation to detect whether the GAN correctly reproduces characteristic clustering patterns, e.g., photopeak and inter-crystal scatter events, practically determining detector performance. For two image patches $\mathbf{x}$ and $\mathbf{y}$ (local regions of the flood map), the local \ac{SSIM} is defined as
\begin{equation}
    \text{SSIM}(\mathbf{x}, \mathbf{y}) = \frac{(2\mu_x\mu_y + C_1)(2\sigma_{xy} + C_2)}{(\mu_x^2 + \mu_y^2 + C_1)(\sigma_x^2 + \sigma_y^2 + C_2)},
\end{equation}
where $\mu_x$ and $\mu_y$ are the local means, $\sigma_x^2$ and $\sigma_y^2$ are the local variances, $\sigma_{xy}$ is the local covariance, and $C_1$, $C_2$ are constants to stabilize the division. The averaged local \ac{SSIM} is computed over sliding windows, based on an uneven number of pixels, across the entire flood map and and averaged to produce a single scalar summary given as the spatial mean of the local SSIM map, ranging from $-1$ (no similarity) to $+1$ (perfect similarity). In the following we refer to this quantity as the \ac{SSIM} for simplicity.\newline
For this evaluation, image resolution plays a crucial role, and in the context of \acp{SiPM}, a meaningful resolution is the typical size of one \ac{SPAD}. State-of-the-art \acp{SiPM} for \ac{PET} have \ac{SPAD} \cite{bronzi_spad_2016} sizes that can range from \qty{10}{\micro \metre} to \qty{100}{\micro \metre}. We therefore utilize an effective flood map resolution of \qty{50}{\micro \metre} by using a sliding window of \numproduct{5x5} pixels combined with a histogram binning of \qty{10}{\micro \metre}.

\paragraph{Bootstrap Confidence Intervals}

Providing uncertainty estimates for visual similarity poses two challenges that prevent the adoption of the approach used in \cref{subsec:dist_sim}. First, the $\gamma$-photon interaction positions along the $Z$-axis are distributed continuously throughout the crystal volume, rather than lying on a regular grid, as in the electron-emission experiments. This precludes the condition-by-condition analysis employed for the \ac{SWS} metric, in which the variance across discrete grid points provides a natural measure of uncertainty. Second, the flood map is an image with strongly non-uniform count statistics (high-count photopeak clusters at crystal centers and lower-count inter-crystal scatter lines).\newline
To obtain statistically well-founded uncertainty estimates under these conditions, we employ bootstrap resampling. For each comparison, we generate $N_\text{bootstrap} = 100$ resampled datasets by randomly drawing (with replacement) from the original \ac{COG} distributions. For each iteration, we recompute the flood map histograms and evaluate the \ac{SSIM}.

\paragraph{Region-stratified and distributional statistics}
Because the spatial mean of the local \ac{SSIM} map is insensitive to deviations limited to only a small fraction of the flood map, we complement it with two additional metrics. First, we report distributional statistics of the local \ac{SSIM} map, specifically the 5th and 1st percentiles of the local \ac{SSIM} distribution, and the fraction of bins whose local \ac{SSIM} falls below a fixed threshold. The threshold is defined as the noise floor of the MC-baseline local \ac{SSIM} distribution.\newline
Second, we report all statistics both aggregated over the full array and stratified by detector region, distinguishing the central crystal ($|X|, |Y| < \qty{1.5}{\milli \metre}$), the eight surrounding outer crystals, and the full array. The regional decomposition enables the identification of spatially localized artifacts that are not resolved in aggregate measures. Bootstrap uncertainties on all region-stratified statistics use the same $N_\text{bootstrap} = 100$ event-level resampling described above.

\subsection{Ablation of Network Modifications}
\label{subsec:ablation}

The two network modifications introduced in \cref{subsec:pi-intro} (Fourier feature encoding with latent mapping network, and unit-norm loss) are incorporated jointly into the final model configuration. To characterize the individual contribution of each modification, we performed a sequential ablation study using three trained networks. The baseline configuration corresponds to the single-crystal optiGAN model architecture and represents the first model. The second model is given as the baseline extended by the Fourier feature encoding and latent mapping network only (optiGAN+Fourier), and the full configuration combining the Fourier encoding with the unit-norm loss term (optiGAN+Fourier+PI) represents the third model. All three networks share a similar parameter count and hyperparameter choices with the single-crystal optiGAN model \cite{trigila_towards_2025}, and were trained on
the identical training set with the same random seed. The results of this ablation, presented in \cref{subsec:ablation_results}, form the basis for selecting the final model configuration used in the evaluation experiments described below.

\paragraph{Ablation Metrics}
The effect of the Fourier feature encoding and latent mapping network is assessed through the training evolution of the condition-wise \ac{SWS} value (see \cref{subsec:dist_sim}) on
the test dataset (fundamental domain, matching the structure of the training dataset), aggregated over all emission conditions and $Z$-layers. To assess where the unit-norm term in the loss function acts most strongly, we evaluate the two converged configurations on the full-array evaluation dataset (see \cref{para:full_array}) and stratify its emission conditions into three regions, being center conditions (located at the geometric center a crystal), boundary conditions (adjacent to an internal crystal-to-crystal interface), and transition-zone conditions (the remaining conditions inside the crystals). For each region, we report the \ac{MAE} of the generated squared momentum norm,
\begin{equation}
\text{MAE} = \frac{1}{N} \sum_{i=1}^{N} \left| \, \| \hat{P}_{g,i} \|_2 - 1 \, \right|,
\end{equation}
together with the fraction of generated samples falling outside a \qty{\pm 5}{\percent} band around the target value of unity. As a complementary spatial view, we also report the \ac{MAE}
stratified by the generated photon detection position on the sensor plane.
 
\subsection{Evaluation Experiments}
\label{subsec:eval_exp}
The generation capabilities of the final model configuration (optiGAN+Fourier+PI) are evaluated using three distinct studies. In each experiment, the model's performance is compared with the ground truth from state-of-the-art GATE10 simulations. To account for the natural fluctuations of Monte Carlo simulations, each evaluation dataset is simulated twice under the exact same simulation settings. Subsequently, the evaluation metrics are applied to compare the two runs and determine the maximum similarity achievable by the \ac{GAN}.

\paragraph{Full Array Evaluation}
\label{para:full_array}
In a first study, the model is evaluated on the full crystal array domain. In this step, we test how the model performs under conditions outside the fundamental domain while maintaining the exact spatial sampling as for the training dataset. The complete dataset incorporates three different $Z$-layers (\qty{1}{\milli \metre}, \qty{5}{\milli \metre}, \qty{9}{\milli \metre}), like the training dataset, while each layer holds 225 electron emission locations covering the full crystal array. At each source location $30$ monoenergetic electrons ($E_{e^-}=\qty{420}{\kilo \eV}$) are emitted. The baseline is provided by the condition-averaged \ac{SWS} value from two Monte Carlo simulations, along with its standard deviation. This baseline is justified because Monte Carlo simulations rely on the same physical models and interaction cross-sections across all conditions, ensuring uniform accuracy regardless of the specific emission location. This assumption cannot be made a priori for the GAN model, as its learned representation may exhibit spatially varying fidelity across different emission conditions. We therefore adopt a more detailed evaluation approach. The \ac{SWS} is computed individually for each emission point, and we subsequently analyze whether the resulting means and standard deviations are consistent across rows and columns of the emission grid for a given $Z$-layer. This allows us to identify potential systematic biases in the model's performance as a function of spatial position within the crystal array.

\paragraph{High-Resolution Evaluation}
\label{para:high_res}
A fundamental distinction between Monte Carlo simulations and learned generative models lies in their spatial generalization behavior. Monte Carlo methods, grounded in first-principles physics, produce accurate results for any emission location within the defined geometry. In contrast, generative models learn distributions from discrete training conditions, raising the question of whether they can reliably interpolate to emission locations not explicitly represented during training.\newline
To investigate this potential limitation, we use a high-resolution dataset probing the model's out-of-distribution generalization capabilities. While training data is acquired on a coarse spatial grid of \qty{0.6}{\milli \metre} step width at a distance from the sensor face of \qty{1}{\milli \metre}, \qty{5}{\milli \metre}, and \qty{9}{\milli \metre} (see \cref{subsec:fundamentalDomain}), the high-resolution dataset uses a much finer step width of \qty{0.25}{\milli \metre} at \qty{1}{\milli \metre}, \qty{3}{\milli \metre}, \qty{5}{\milli \metre}, \qty{7}{\milli \metre}, and \qty{9}{\milli \metre} height. At each source location $30$ monoenergetic electrons ($E_{e^-}=\qty{420}{\kilo \eV}$) are emitted. This experimental design enables systematic assessment of whether the model has learned a continuous, physically meaningful mapping from emission coordinates to optical photon distributions, or whether its performance degrades for conditions lying between training positions. Such interpolation capability is essential for practical applications where arbitrary interaction locations must be modeled accurately.\newline
Because high-resolution sampling significantly increases the number of emission locations and, consequently, the simulation time and storage requirements, we limit ourselves to emission locations within the fundamental domain. Furthermore, sources located exactly at the crystal boundaries ($\{ (x,y): x = \qty{1.5}{\milli \metre} \lor y = \qty{1.5}{\milli \metre}\}$) are excluded from the evaluation dataset, as the physical behavior at these interfaces is mathematically ill-defined for both the Monte Carlo simulation as well as the \ac{GAN}.\newline
Following the evaluation methodology established in the full array study, we compute the \ac{SWS} individually for each emission condition and analyze spatial consistency across the grid, considering the row and column mean values and standard deviations. For conditions that are located in a row or column having fewer than four other conditions, the row or column-averaged standard deviation is used.

\paragraph{Pencil Beam Evaluation}
\label{para:pencil_beam}
While electron emission within the crystal array provides a well-controlled setting for generating high-fidelity training and evaluation datasets, a more realistic scenario arises when annihilation photons are created outside the detection volume and impinge on the array's surface. Besides that, often in experimental PET detector characterization, the primary quantity of interest is not the detailed optical photon distribution itself, but rather the resulting flood map typically computed via \ac{COG} estimation from the sensor signals.\newline
To evaluate the model's performance in this practical use case, we apply it to the pencil-beam irradiation dataset. The data are created by irradiating the center of each of the nine crystals with $10^4$ monoenergetic $\gamma$-photons ($E_{\gamma} = \qty{511}{\kilo \eV}$) being located \qty{1}{\milli \metre} above the array's surface. For each recorded $\gamma$-interaction, the associated optical photon data are clustered and stored. Subsequently, the \ac{GAN} is conditioned on the $\gamma$-photons interaction position inside the array, and the corresponding optical photon data are generated.\newline

\begin{figure}[htb]
\centering
\includegraphics[width=0.9\linewidth]{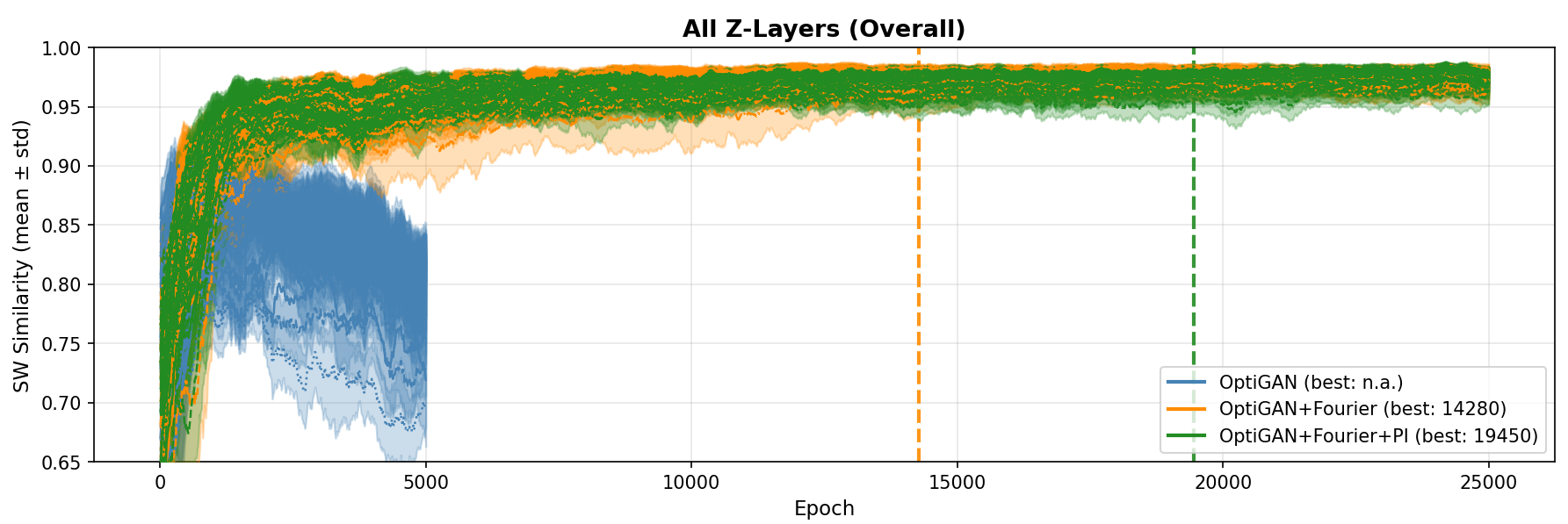}
\caption{Condition-wise \ac{SWS} during training for the three configurations, aggregated over all conditions. Vertical dashed lines indicate the best-performing epoch for the two converged configurations. The unmodified optiGAN configuration was terminated at 5000 epochs due to a lack of convergence.}
\label{fig:SWS-Comp}
\end{figure}

\section{Results}

\subsection{Ablation of Network Modifications}
\label{subsec:ablation_results}

The results of the ablation study described in \cref{subsec:ablation} are presented below. The training evolution of the condition-wise \ac{SWS} characterizes the effect of the Fourier feature encoding, and the region-stratified analysis of the generated momentum-norm error characterizes the effect of the unit-norm loss term.

\paragraph{Fourier Feature Encoding and Latent Mapping Network}
\Cref{fig:SWS-Comp} shows the condition-wise \ac{SWS} during training under identical conditions for the unmodified single-crystal optiGAN architecture, optiGAN extended by the Fourier feature encoding together with the latent mapping network (optiGAN+Fourier), and the full proposed model including the unit-norm loss term (optiGAN+Fourier+PI).\newline
The original architecture did not reach a stable training regime on the \numproduct{3 x 3} array within 5,000 epochs and was terminated due to the absence of any convergence trend and an increasing fraction of conditions with collapsed \ac{SWS} values. Both optiGAN+Fourier and optiGAN+Fourier+PI reached stable plateaus at \ac{SWS} values of approximately 0.95-0.98 from 10,000 epochs onward.
This observation indicates that the latent-space augmentation enables stable training on the array geometry while preserving the generator and discriminator architectures of the single-crystal.

\paragraph{Unit-Norm Loss Term}
We compare optiGAN+Fourier and optiGAN+Fourier+PI at their respective best-epoch checkpoints (\cref{fig:SWS-Comp}). Values of the momentum-norm \ac{MAE} and the tail fraction outside a \qty{\pm 5}{\percent} band around unity, aggregated across all three $Z$-layers of the full-array evaluation dataset, are listed in \cref{tab:PI-Loss}. The reduction induced by the unit-norm term is most pronounced for boundary-adjacent conditions, while transition-zone conditions remain essentially unchanged in \ac{MAE}. After training with the term, all three regions reach comparable tail-fraction values.

\begin{table}[H]
\caption{\ac{MAE} of the squared momentum norm and fraction of generated samples outside a \qty{\pm 5}{\percent} band around unity. Emission conditions are stratified by position type relative to crystal geometry, pooled across all nine crystals.}
\begin{tabular}{@{}ccccc@{}}
\toprule
\multirow{2}{*}{Condition region} & \multicolumn{2}{c}{MAE} & \multicolumn{2}{c}{Tail Fraction (\qty{\pm 5}{\percent})} \\ \cmidrule(l){2-5}
 & optiGAN+Fourier & optiGAN+Fourier+PI & optiGAN+Fourier & optiGAN+Fourier+PI \\ \midrule
Crystal center           & 0.02537          & 0.02263          & \qty{11.71}{\percent}          & \qty{7.81}{\percent}          \\
Inter-crystal boundary & 0.02587 & 0.02275 & \qty{12.33}{\percent} & \qty{7.79}{\percent} \\
Transition zone  & 0.02447          & 0.02420          & \qty{10.76}{\percent}          & \qty{8.22}{\percent}          \\ \bottomrule
\end{tabular}
\label{tab:PI-Loss}
\end{table}

\begin{figure}[H]
\centering
\includegraphics[width=0.7\linewidth]{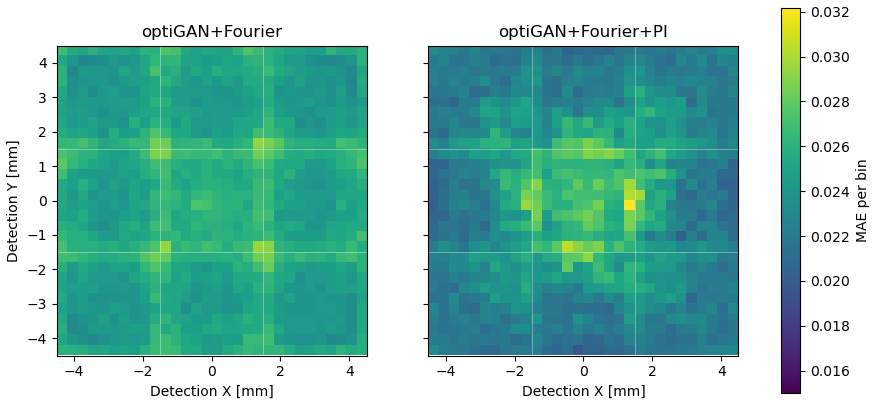}
\caption{\ac{MAE} of the squared momentum norm stratified by the photon detection position on the sensor plane. The left image shows the configuration optiGAN+Fourier and the right image displays optiGAN+Fourier+PI.}
\label{fig:MAE-Comp}
\end{figure}

A complementary spatial view is shown in \cref{fig:MAE-Comp}. Without the unit-norm term (left), elevated \ac{MAE} values follow the inter-crystal interfaces of the \numproduct{3x3} array, aligned with the geometry of the crystal boundaries. With the term added (right), this boundary structure is reduced.

\paragraph{Selection of the Final Model Configuration}
Based on this ablation study, we selected the full configuration combining Fourier feature encoding with the unit-norm loss term (optiGAN+Fourier+PI) for subsequent experiments. The Fourier feature encoding and latent mapping network enables stable training on the array data, whereas the unit-norm loss term addresses the challenge of modeling of photon transport near inter-crystal interfaces and neighboring crystals, which is geometrically complex and only arise in arrays. The interplay of both contributions in the final model and the degradation observed in the flood-map fidelity in the central crystal, is examined in \cref{subsec:penBeamEva}.

\begin{figure}[htb]
\centering
\subfloat[\ac{SWS} values for conditions located \qty{1}{\milli \metre} above the sensor.]{%
\includegraphics[width=0.31\textwidth]{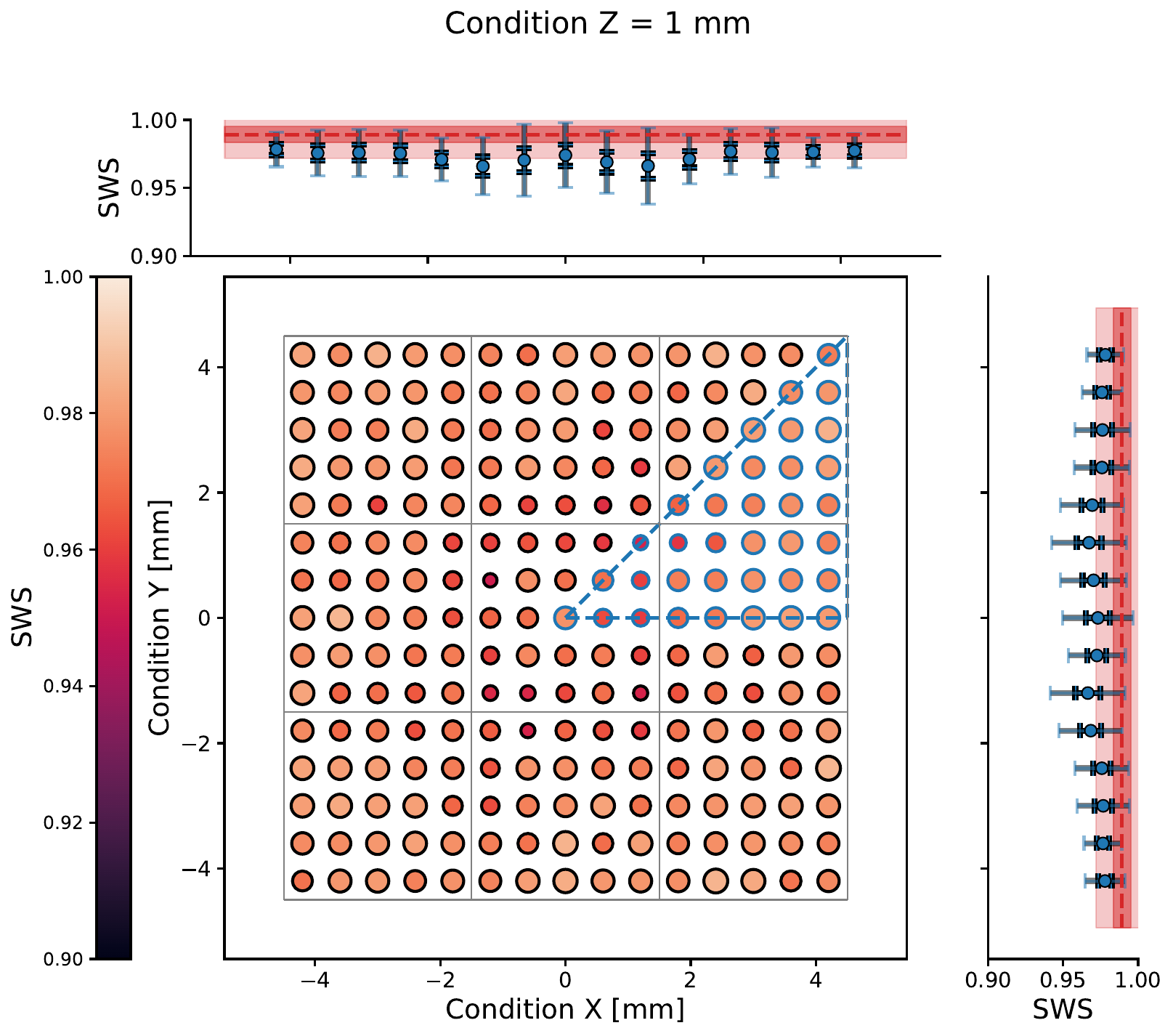}
\label{fig:FA_analysis1}}
\quad
\subfloat[\ac{SWS} values for conditions located \qty{5}{\milli \metre} above the sensor.]{%
\includegraphics[width=0.31\textwidth]{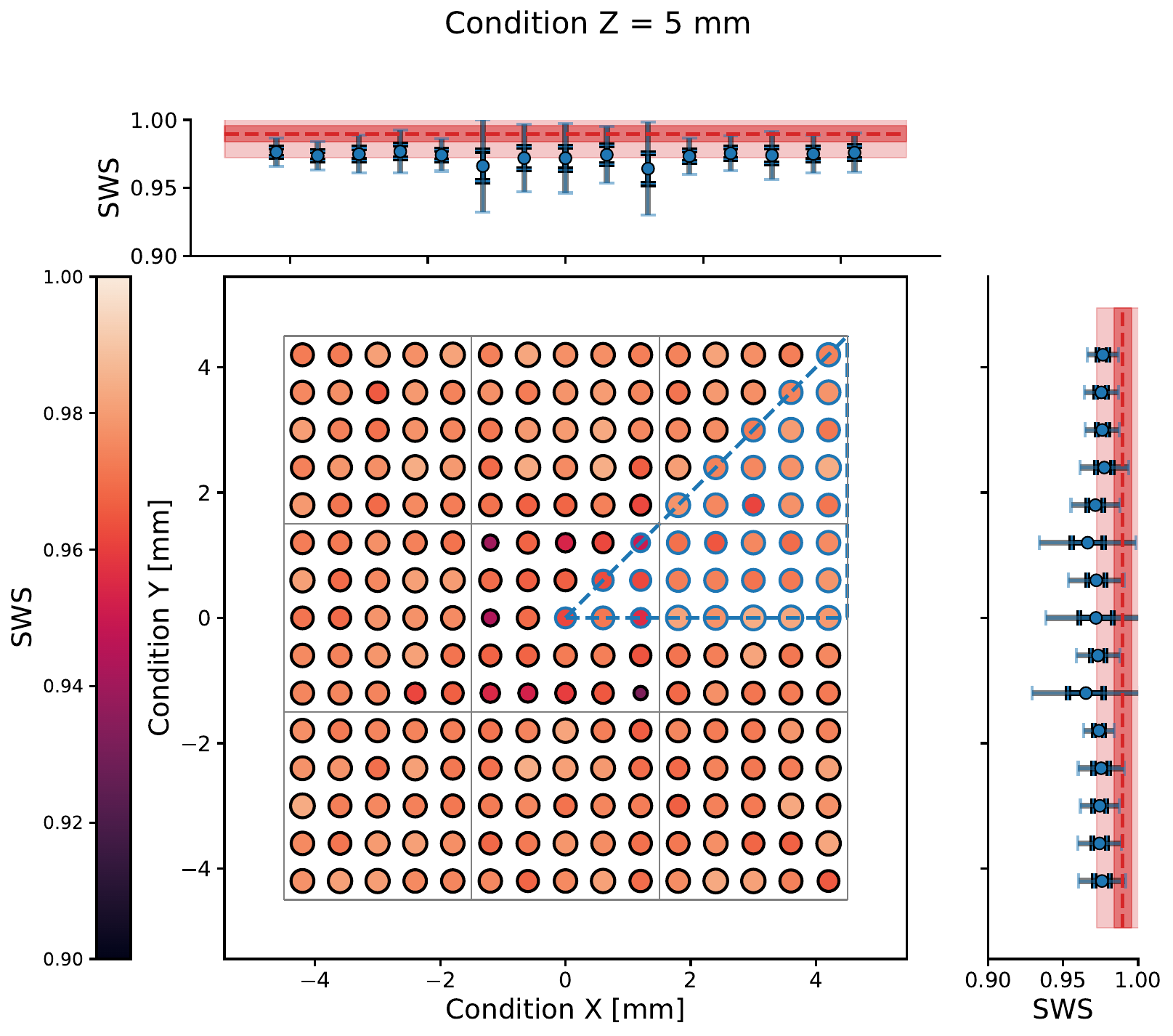}
\label{fig:FA_analysis5}}
\quad
\subfloat[\ac{SWS} values for conditions located \qty{9}{\milli \metre} above the sensor.]{%
\includegraphics[width=0.31\textwidth]{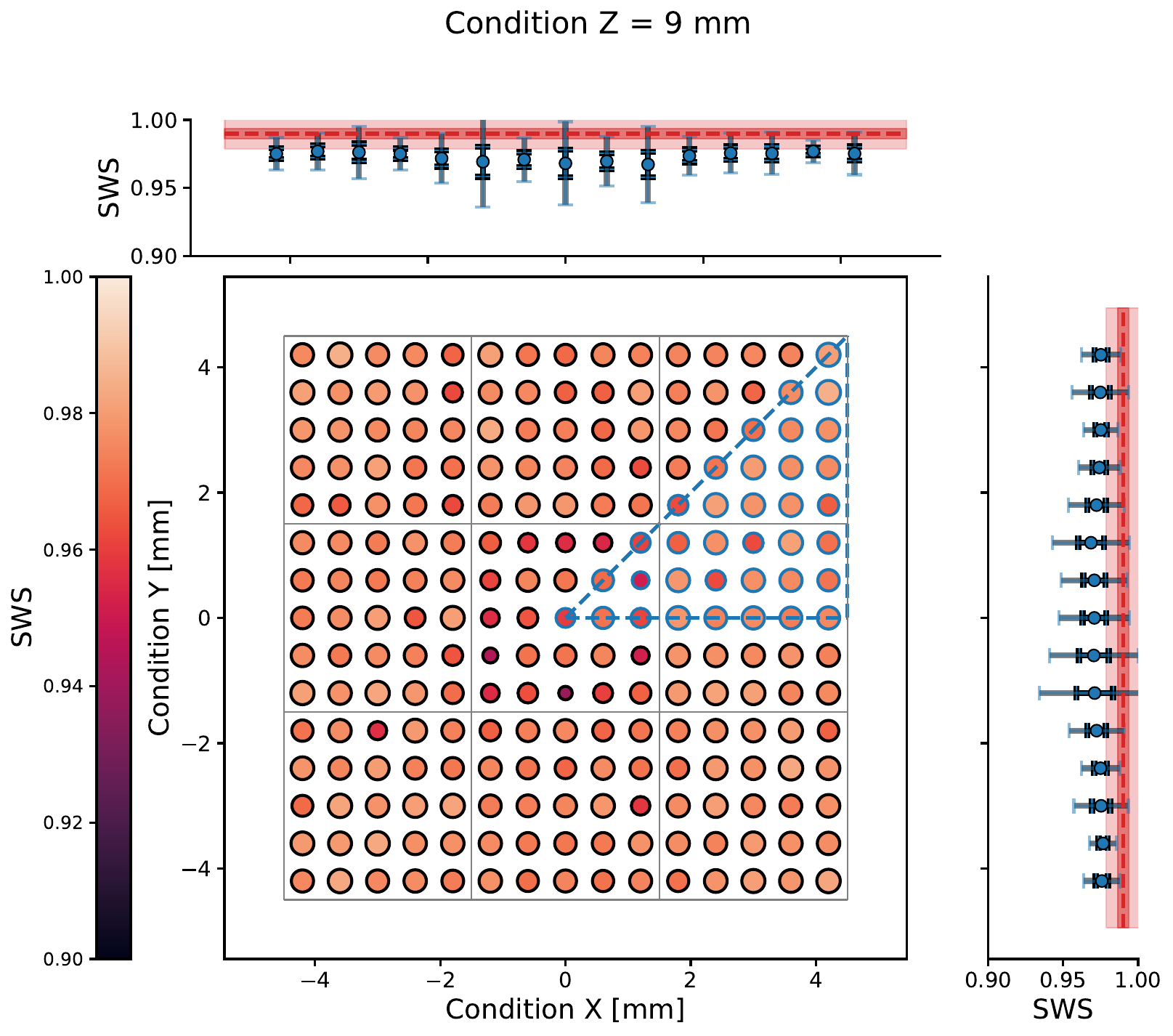}
\label{fig:FA_analysis9}}
\caption{\ac{SWS} results for the full array evaluation. The location of each scatter point represents the condition location, while the color visualizes the estimated \ac{SWS} value. The row and column-wise results are depicted in the side and top plots. The blue point indicates the mean, and the bright and transparent error bars show one and three standard deviations. The dashed red line, with a bright and transparent red band, represents the baseline \ac{SWS} performance (mean, one standard deviation, three standard deviations) achieved by two independent simulation runs with the same settings. Points with a blue edge color coincide with emission locations that were originally present during the training.}
\label{fig:FA_analysis}
\end{figure}

\subsection{Full Array Evaluation}
\label{subsec:ana_FA}

The results of the full array evaluation are depicted in \cref{fig:FA_analysis}. All \ac{SWS} row and column values agree with the baseline Monte Carlo simulation at least within a $3 \sigma$-range (shown by the pink region around the red dashed line). Conditions outside the fundamental domain show the same similarity performance as conditions inside the fundamental domain. Furthermore, \ac{SWS} values of conditions located in the lowest $Z$-layer ($Z=\qty{1}{\milli \metre}$) and central crystal are slightly decreased compared to the values of the side and corner crystals. This effect is only partially observed for higher $Z$-layers, where mostly only the conditions at the neighboring crystal boundaries show decreased values.

\subsection{High-Resolution Evaluation}
\begin{figure}[htb]
\centering
\subfloat[\ac{SWS} values for conditions located \qty{1}{\milli \metre} above the sensor.]{%
\includegraphics[width=0.31\textwidth]{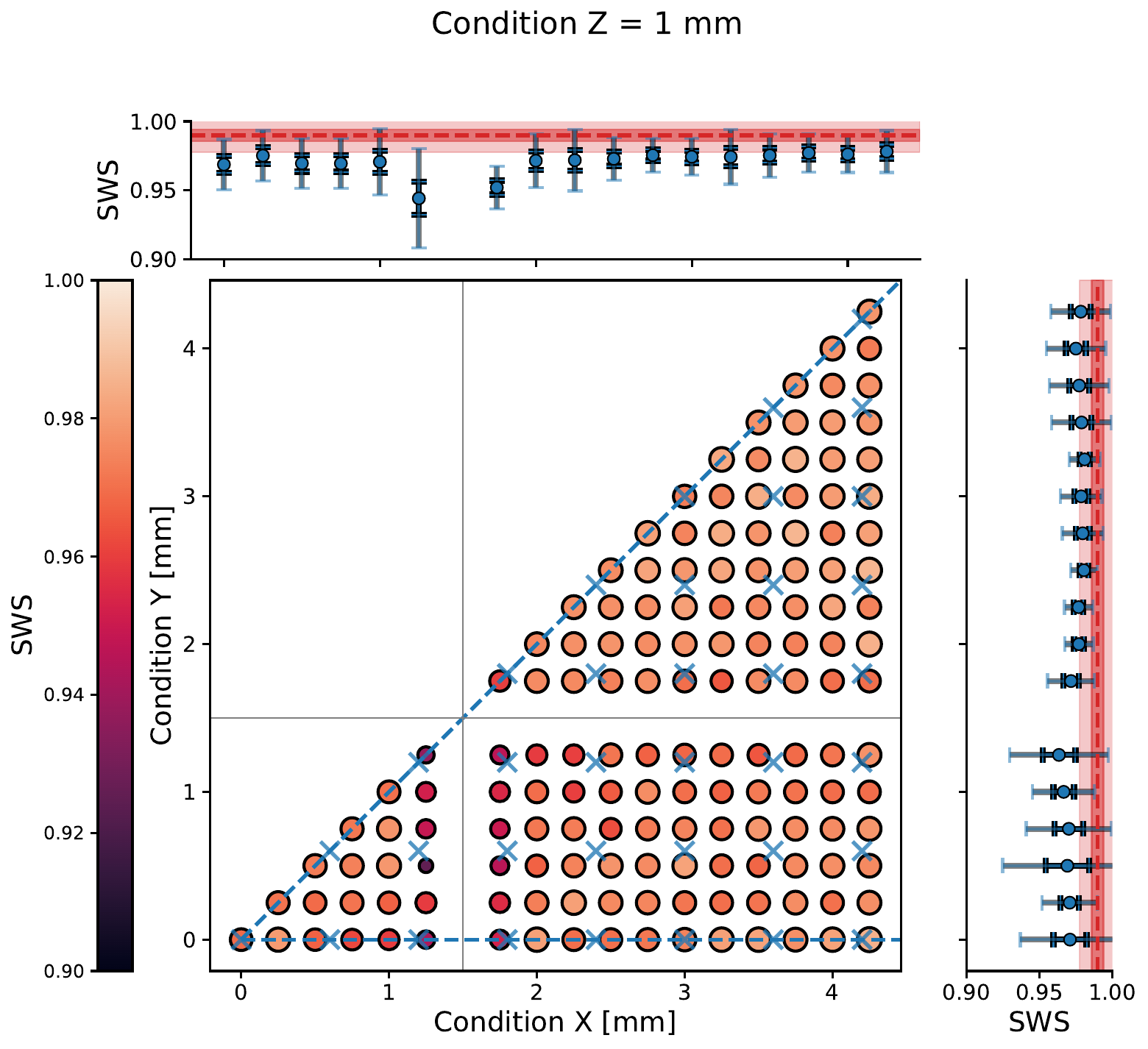}}
\quad
\subfloat[\ac{SWS} values for conditions located \qty{3}{\milli \metre} above the sensor.]{%
\includegraphics[width=0.31\textwidth]{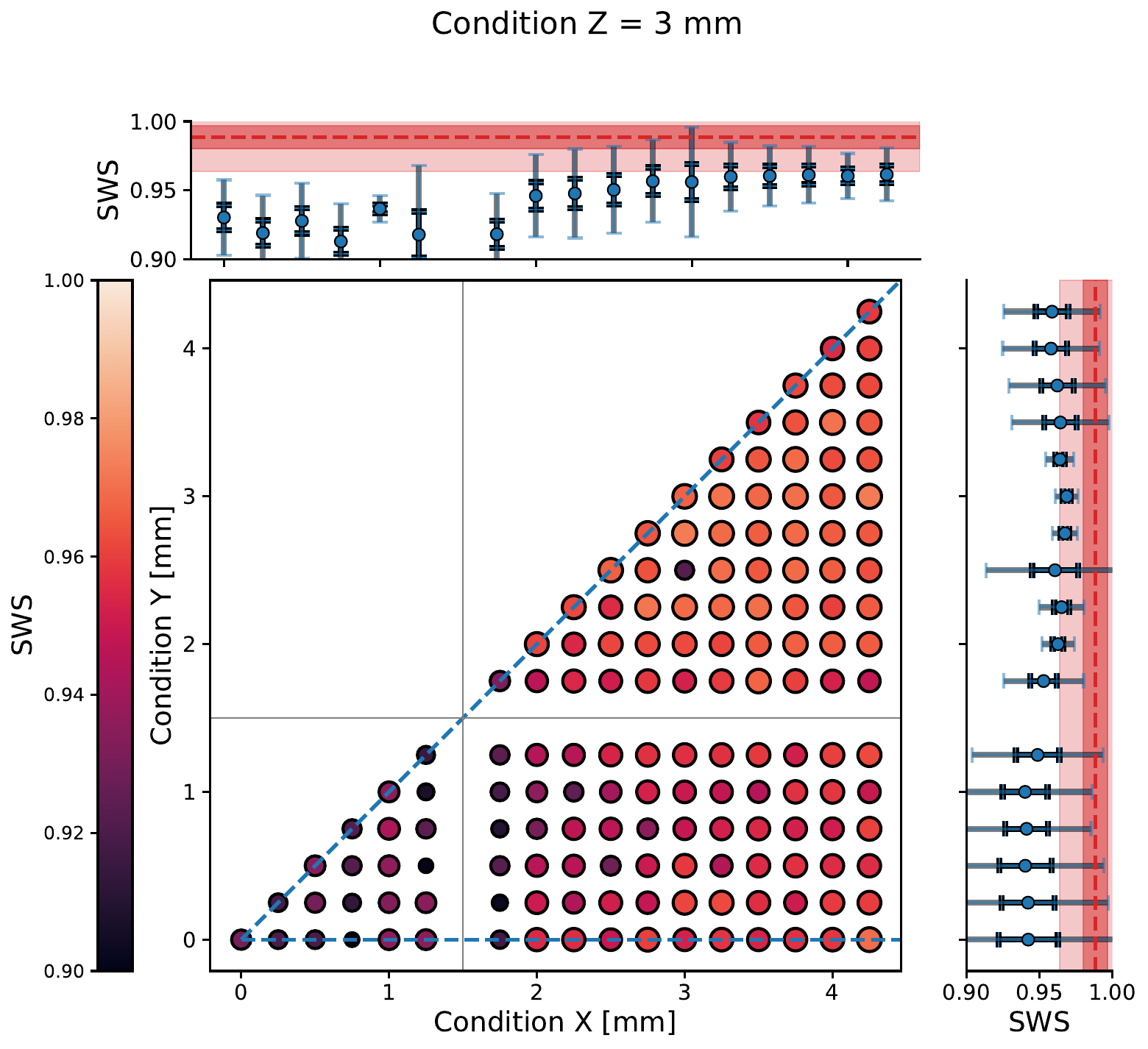}}
\quad
\subfloat[\ac{SWS} values for conditions located \qty{5}{\milli \metre} above the sensor.]{%
\includegraphics[width=0.31\textwidth]{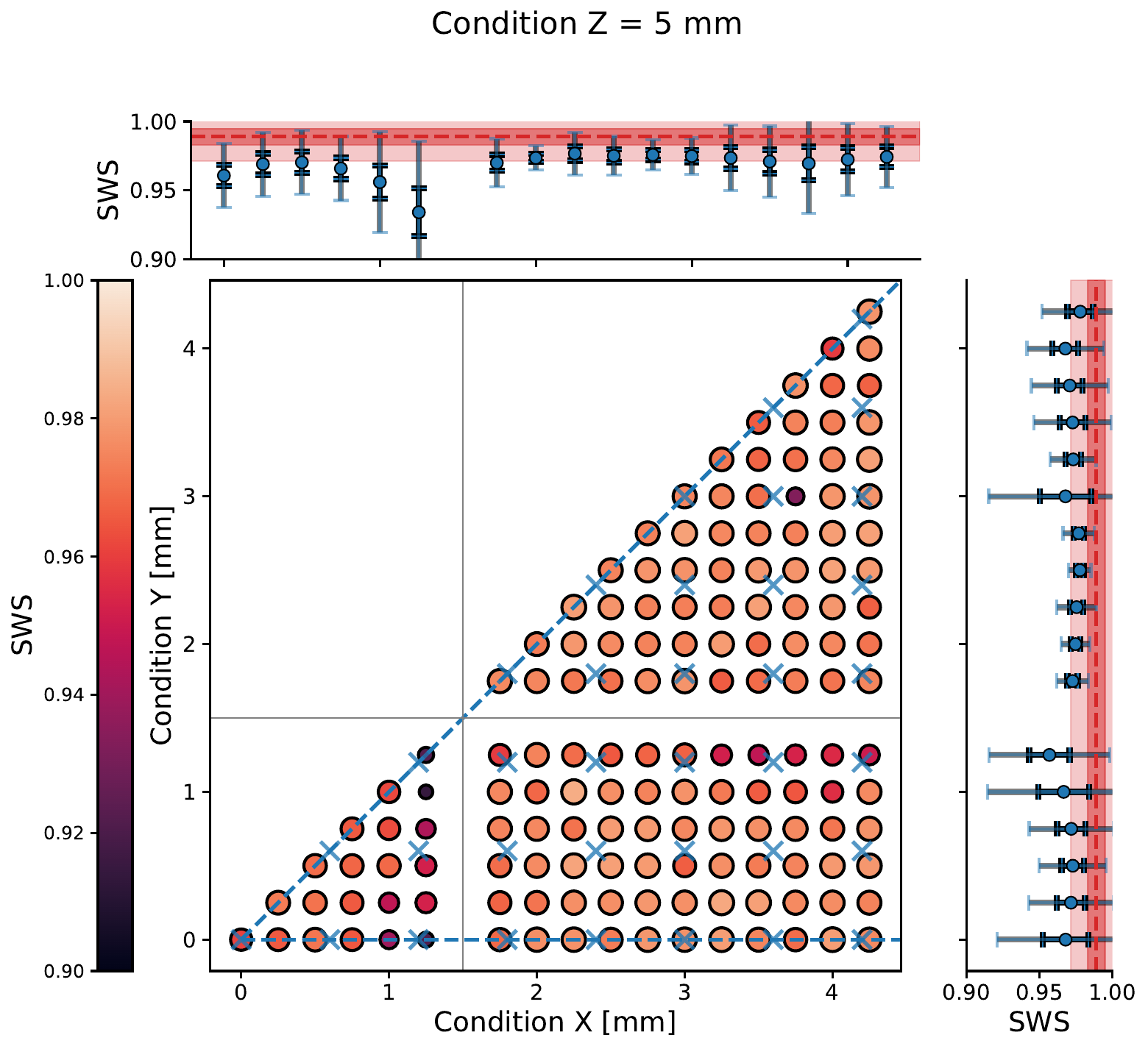}}
\quad
\subfloat[\ac{SWS} values for conditions located \qty{7}{\milli \metre} above the sensor.]{%
\includegraphics[width=0.31\textwidth]{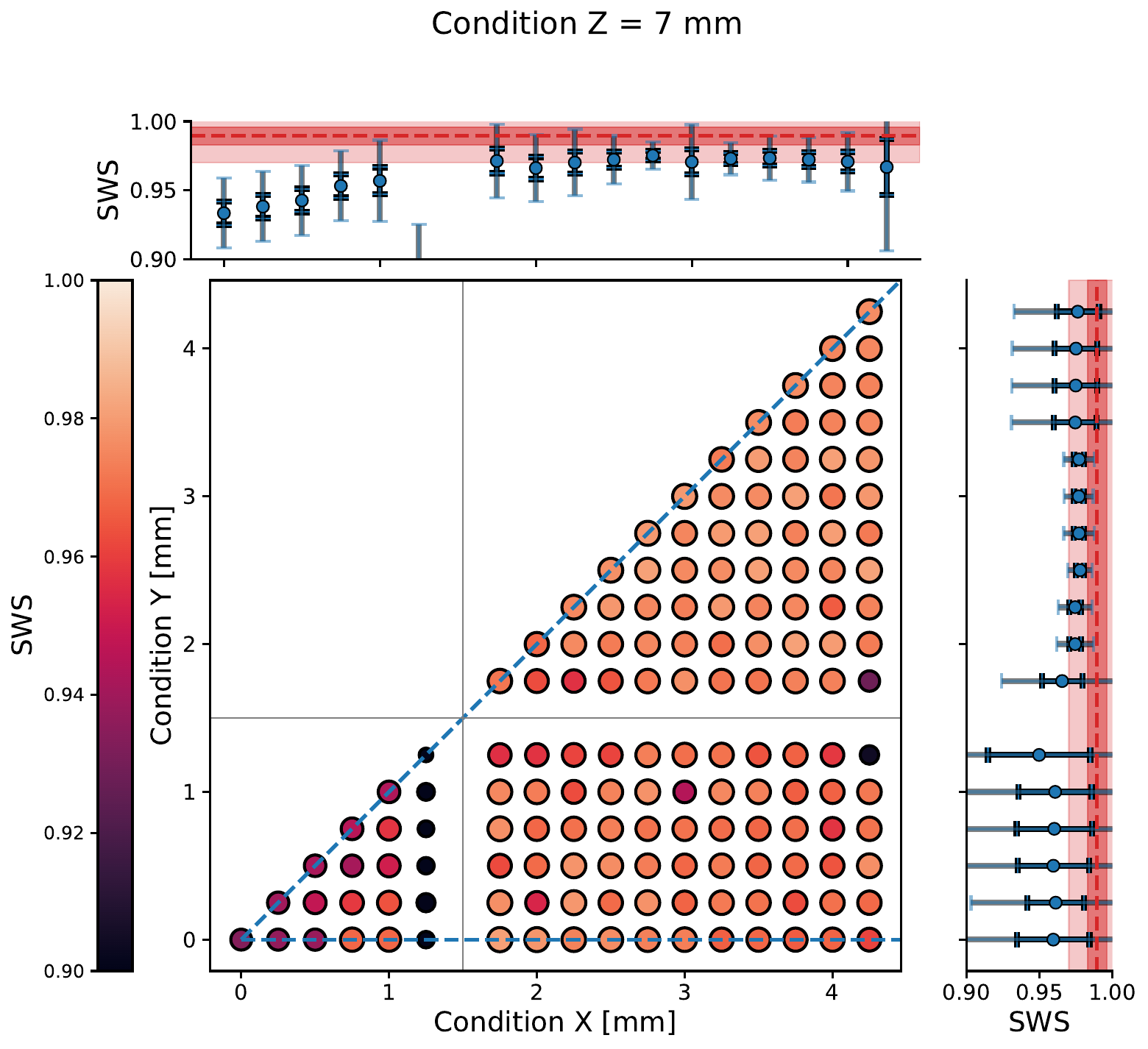}}
\quad
\subfloat[\ac{SWS} values for conditions located \qty{9}{\milli \metre} above the sensor.]{%
\includegraphics[width=0.31\textwidth]{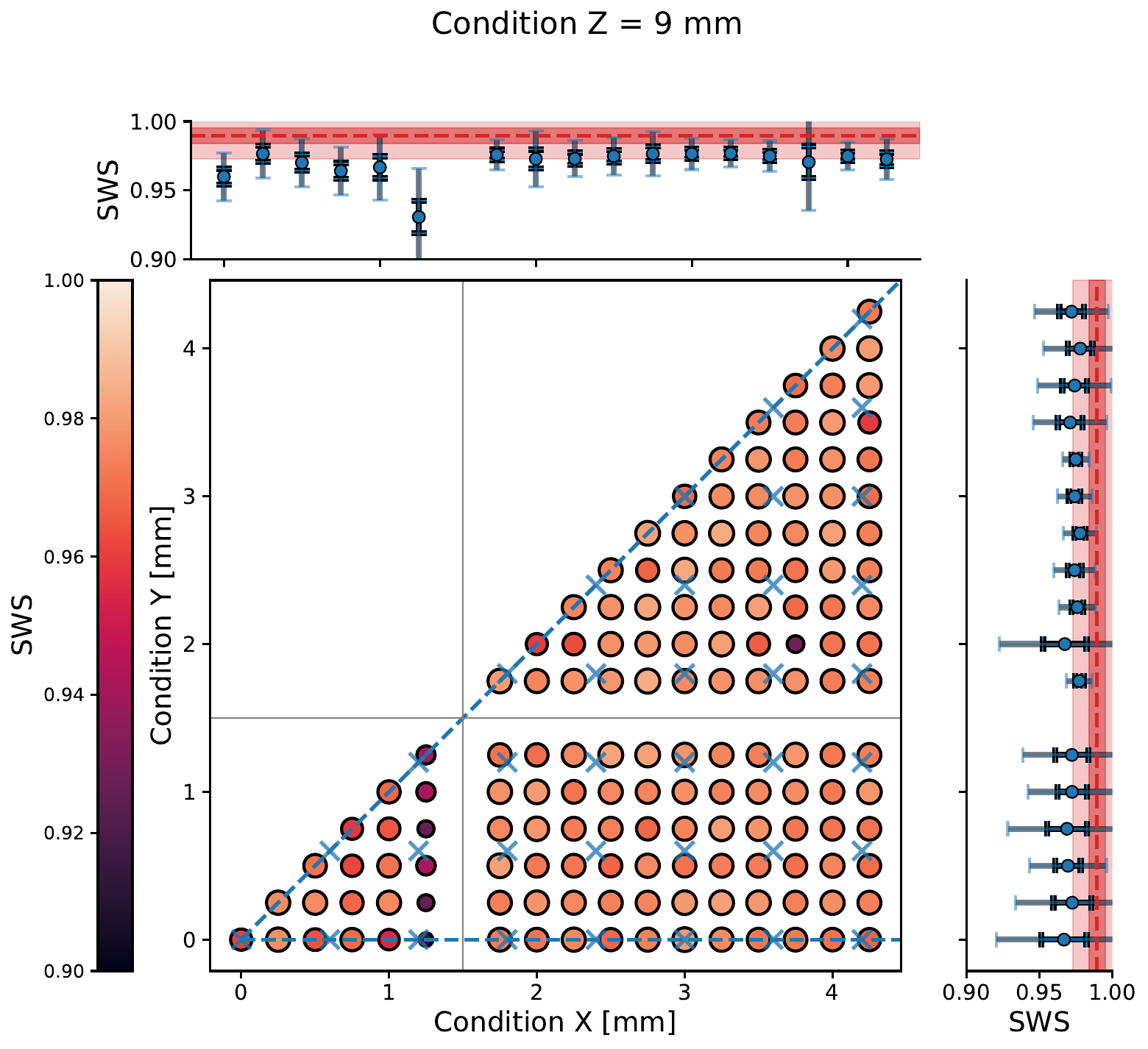}}
\caption{\ac{SWS} results for the high-resolution evaluation. The location of each scatter point represents the condition location, while the color corresponds to the estimated \ac{SWS} value. The row and column profiles are shown in the side and top plots. The blue point indicates the mean, and the bright and transparent error bars show one and three standard deviations. For conditions that are located in a row or column having fewer than four other conditions, the row or column-averaged standard deviation is used. The dashed red line, with a bright and transparent red band, represents the baseline \ac{SWS} performance (mean, one standard deviation, three standard deviations) achieved by two independent simulation runs with the same settings. Transparent blue crosses ($Z = \qty{1}{\milli \metre}, \qty{5}{\milli \metre}, \qty{9}{\milli \metre}$) mark the emission locations that were originally present during training.}
\label{fig:HR_analysis}
\end{figure}

The results of the high-resolution evaluation are depicted in \cref{fig:HR_analysis}. Considering the top and side plot, \qty{85}{\percent} of the column and \qty{100}{\percent} of the row \ac{SWS} values agree with the Monte Carlo ground truth within $3\sigma$. For a given \ac{DOI}-layer, no systematic performance degradation is observed for evaluation conditions located further from training conditions (blue crosses). Points lying midway between training grid positions achieve comparable \ac{SWS} values to those in immediate proximity of training data. The best performance is observed for the lowest, middle, and highest $Z$-layer, while the layers in between show a slightly decreased similarity. Similar to the previous evaluation, \cref{subsec:ana_FA}, \ac{SWS} values of conditions located in the central crystal are lower than their counterparts within the side or edge crystal. Besides that, the plots reveal that conditions near the crystal boundary running parallel to the $x$-axis show better performance than conditions located along the crystal boundary parallel to the $y$-axis. \Cref{tab:doi_layerwise_sws} summarizes the layer-wise \ac{SWS} statistics. Pairwise $\sigma$-distances between the three known $Z$-layers range between \num{0.1} and \num{0.3}, indicating excellent agreement between those layers. The two unknown layers likewise differ by $0.4\sigma$ from each other and show an increased inter-layer variability. Comparing the set of unknown $Z$-layers with the set of known layers results in a deviation of $2.0\sigma$ under assumption of the intrinsic variability of the known layer.

\begin{table}[htb]
\centering
\caption{Row and column marginal \ac{SWS} statistics per \ac{DOI} layer for optiGAN+Fourier+PI on the high-resolution evaluation dataset. Values are rounded such that the uncertainty carries one significant figure. The rightmost column reports the $\sigma$-distance of each layer to the trained-layer pool ($\mu_\mathrm{trained} = 0.971$, $\sigma_\mathrm{trained} = 0.008$, $n = 102$), computed as $|\mu_\mathrm{layer} - \mu_\mathrm{trained}| / \sigma_\mathrm{trained}$.}
\begin{tabular}{lcccc}
\toprule
$Z$-layer [mm] & Type      & $\mu_\mathrm{SWS}$ & $\sigma_\mathrm{SWS}$ & $d_\sigma$ vs.\ trained pool \\
\midrule
1 & known   & 0.972 & 0.008 & 0.1 \\
3 & unknown & 0.95  & 0.02  & 2.6 \\
5 & known   & 0.970 & 0.008 & 0.1 \\
7 & unknown & 0.96  & 0.02  & 1.4 \\
9 & known   & 0.971 & 0.008 & 0.0 \\
\bottomrule
\end{tabular}
\label{tab:doi_layerwise_sws}
\end{table}

\subsection{Pencil Beam Evaluation}
\label{subsec:penBeamEva}
\begin{figure}[htb]
\centering
\subfloat[\ac{COG} distribution associated with baseline Monte-Carlo simulation and corresponding \ac{SSIM} value.]{%
\includegraphics[width=0.45\textwidth]{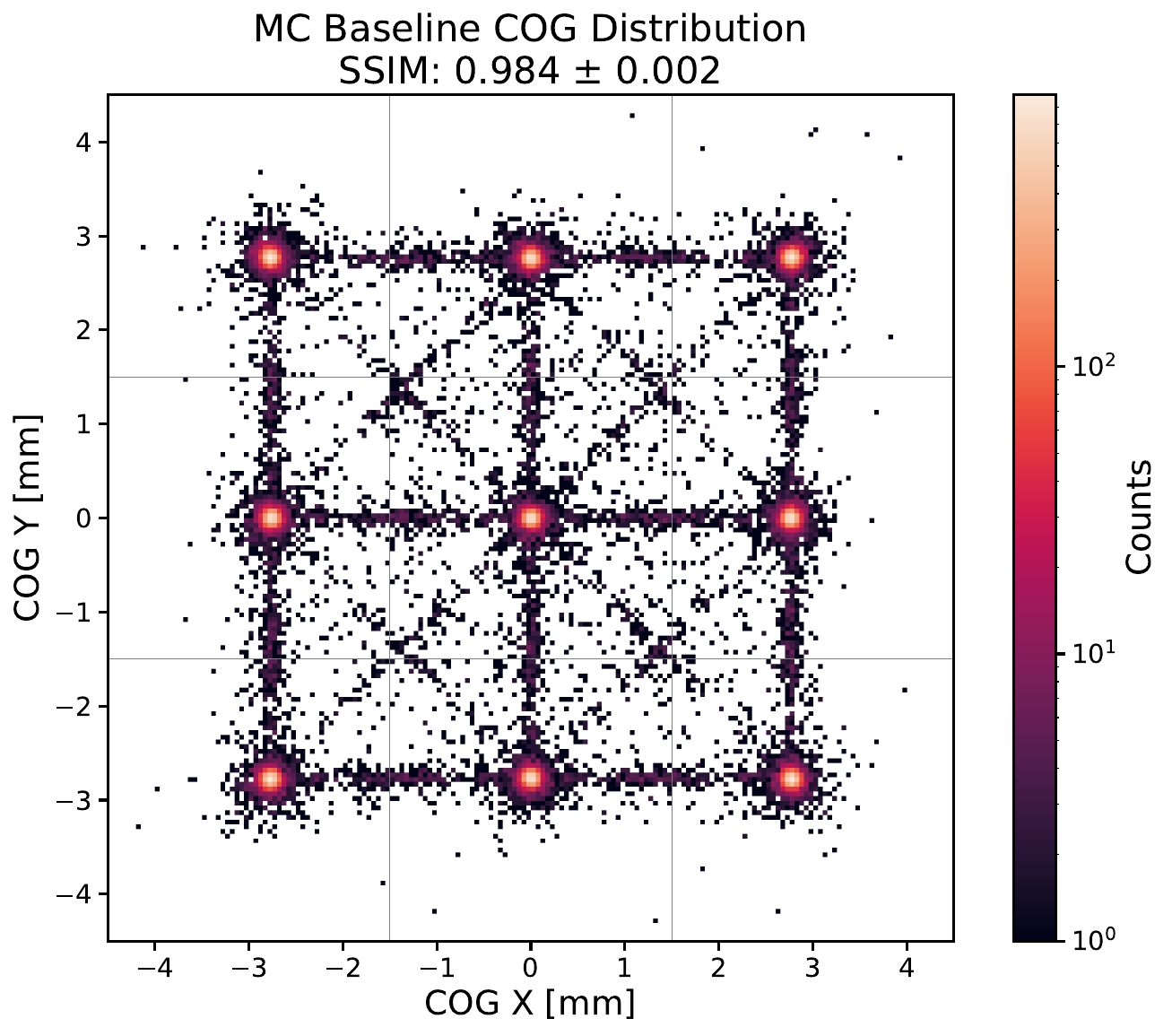}
}
\quad
\subfloat[\ac{COG} distribution associated with optiGAN+Fourier+PI and corresponding \ac{SSIM} value.]{%
\includegraphics[width=0.45\textwidth]{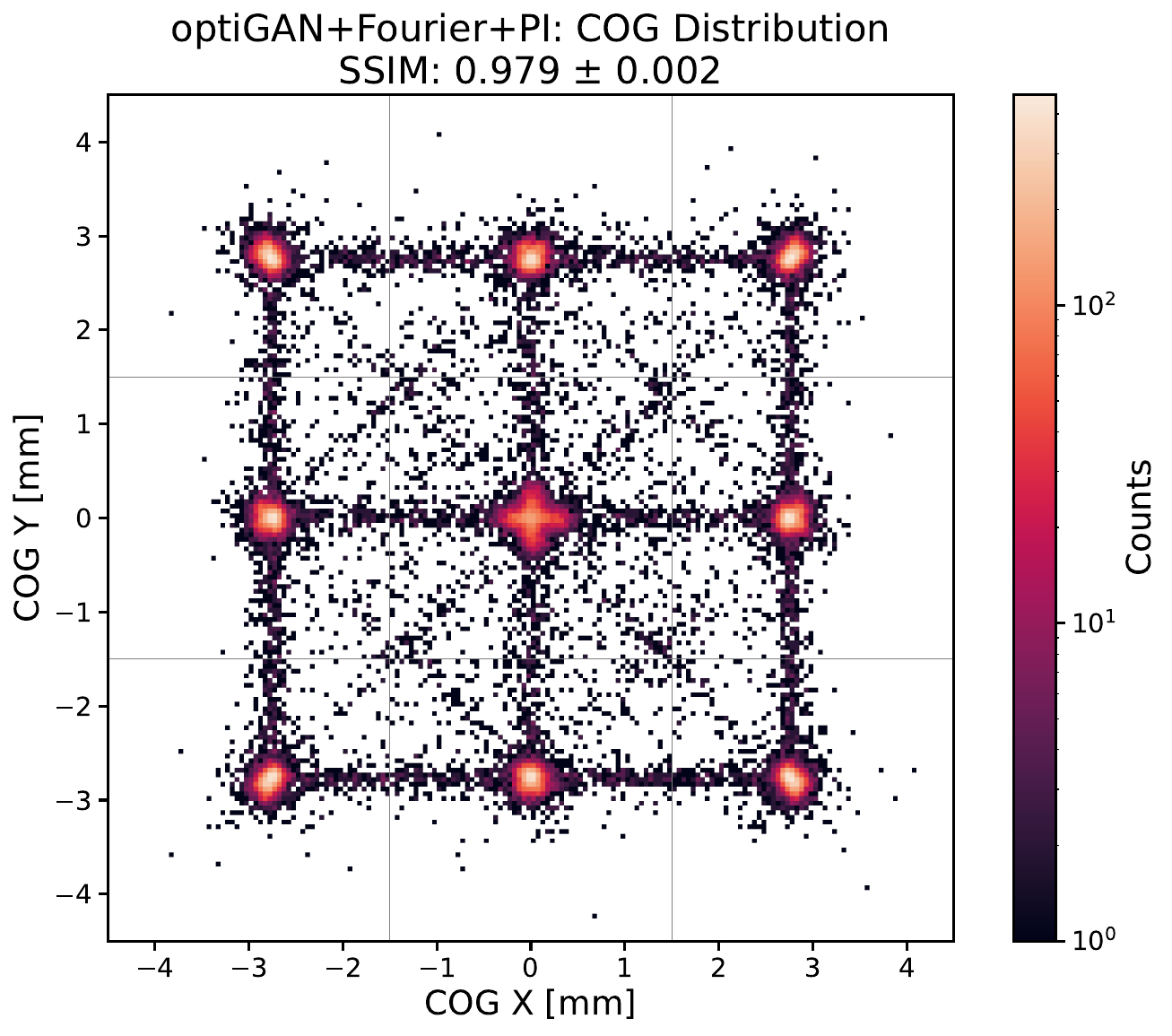}
}
\caption{\ac{COG} distribution generated by the baseline Monte-Carlo simulation and optiGAN+Fourier+PI. Photopeak (bright clusters) and inter-scatter (lines) events are clearly visible in both figures. The gray grid represents the crystals of the detector array.}
\label{fig:PB_COGdist}
\end{figure}

The flood map based on the \ac{GAN}-generated optical photon data and Monte Carlo simulation data is shown at \cref{fig:PB_COGdist} with the estimated mean local \ac{SSIM} value. Both \ac{SSIM} values are within a $3\sigma$-agreement, with the baseline Monte Carlo simulation achieving $\num{0.984 \pm 0.002}$, and $\num{0.979 \pm 0.002}$. The \ac{GAN} is able to reproduce photopeak events as well as inter-crystal scatter events. A dominant artifact can be seen in the photopeak cluster of the central crystal, showing a cross-like structure, which is not observed in the Monte Carlo ground truth.\newline
The ablation analysis in \cref{subsec:ablation_results} already showed that the unit-norm loss term reshapes the spatial distribution of the model's momentum-norm error, with elevated deviations emerging in the center of the sensor plane (see \cref{fig:MAE-Comp}). This observation concerns the momentum outputs, for which a per-sample ground truth exists ($\|\hat{\mathbf{P}}\|^2 = 1$). However, the flood map is computed from the generated energies and detection coordinates, so momentum deviations cannot contribute. Furthermore, no analogous per-sample ground truth exists for the generated detection coordinates $(X, Y)$, since the Monte Carlo reference is a distribution of $(X, Y)$ per emission condition, not an expected value for individual samples. Any assessment of the loss term's effect on the flood map fidelity must therefore be performed at the distribution level, per emission condition. We compare the final model (optiGAN+Fourier+PI) with the ablation configuration without the unit-norm loss term (optiGAN+Fourier) using local \ac{SSIM} stratified by emitting crystal, which resolves the per-condition flood-map fidelity for the central and outer crystals separately. 

\begin{figure}[htb]
\centering
\subfloat[\ac{COG} distribution associated with baseline Monte-Carlo simulation and corresponding \ac{SSIM} value.]{%
\includegraphics[width=0.45\textwidth]{MC_baseline_COG_distribution.pdf}
}
\quad
\subfloat[\ac{COG} distribution associated with optiGAN+Fourier and corresponding \ac{SSIM} value.]{%
\includegraphics[width=0.45\textwidth]{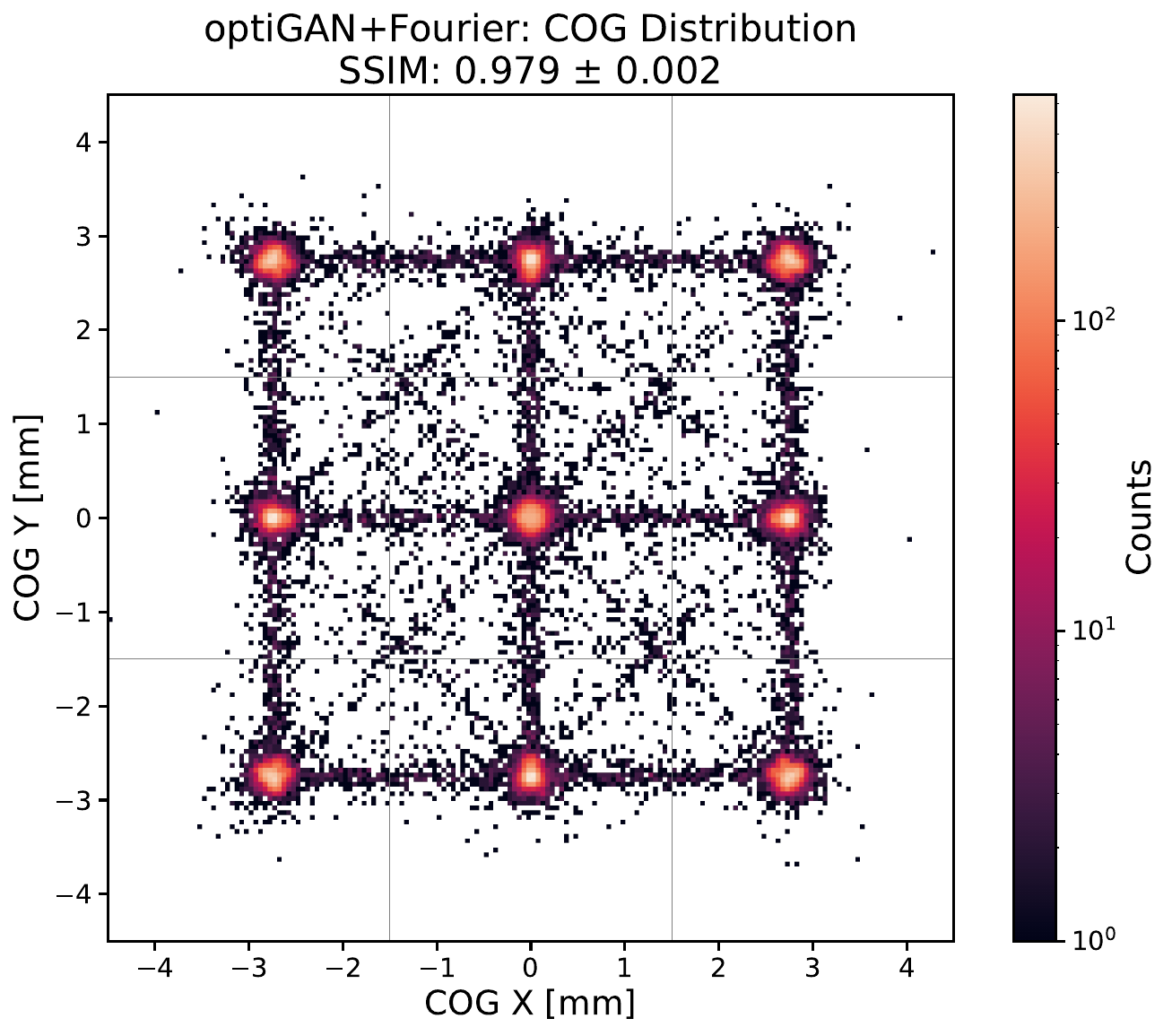}
}
\caption{\ac{COG} distribution generated by the baseline Monte-Carlo simulation and optiGAN+Fourier (Left and right, respectively).}
\label{fig:attribution_floodmap}
\end{figure}

\Cref{fig:attribution_floodmap} shows the resulting flood map for the configuration trained without the unit-norm loss together with the Monte Carlo reference. The cross-shaped artifact is absent in the data generated by this model configuration. Furthermore, the visual comparison shows that the photopeak clusters generated by optiGAN+Fourier appear slightly bigger in size, especially in the corner crystals. The vertical and horizontal lines, which occur due to light-sharing and inter-crystal scatter, appear more defined for the configuration employing the unit-norm loss. Averaged over the full array, the mean local \ac{SSIM} is statistically indistinguishable between the two configurations.

\Cref{tab:attribution_ssim} reports the region-stratified local \ac{SSIM} statistics introduced in \cref{subsec:visual_sim} for both configurations against the Monte Carlo reference.

\begin{table}[H]
\centering
\caption{Region-stratified local \ac{SSIM} statistics for optiGAN+Fourier and optiGAN+Fourier+PI, together with the Monte Carlo baseline. Values are point estimates $\pm$ bootstrap standard deviation ($N_\text{bootstrap} = 100$), rounded such that the
uncertainty carries one significant figure. $\text{Pct}_{<\tau}$ denotes the fraction of bins in the local \ac{SSIM} map falling below the threshold $\tau = 0.193$, defined as the noise floor of the MC-baseline local \ac{SSIM} distribution. $\text{P5}$ and $\text{P1}$ are the 5th and 1st percentiles of the local \ac{SSIM} distribution.}
\begin{tabular}{@{}llccc@{}}
\toprule
Region & Metric & optiGAN+Fourier & optiGAN+Fourier+PI & MC baseline \\ \midrule
\multirow{4}{*}{Full array}
    & Mean \ac{SSIM}             & $0.979 \pm 0.002$ & $0.979 \pm 0.002$ & $0.984 \pm 0.002$ \\
    & $\text{Pct}_{<\tau}$ [\%]  & $0.33 \pm 0.04$   & $0.17 \pm 0.05$   & $0.000 \pm 0.004$ \\
    & $\text{P5}$                & $0.91 \pm 0.02$   & $0.92 \pm 0.02$   & $0.92 \pm 0.01$   \\
    & $\text{P1}$                & $0.50 \pm 0.02$   & $0.49 \pm 0.02$   & $0.73 \pm 0.02$   \\ \midrule
\multirow{4}{*}{Outer 8 crystals}
    & Mean \ac{SSIM}             & $0.980 \pm 0.002$ & $0.982 \pm 0.002$ & $0.986 \pm 0.002$ \\
    & $\text{Pct}_{<\tau}$ [\%]  & $0.32 \pm 0.04$   & $0.09 \pm 0.04$   & $0.000 \pm 0.005$ \\
    & $\text{P5}$                & $0.92 \pm 0.01$   & $0.92 \pm 0.02$   & $0.92 \pm 0.01$   \\
    & $\text{P1}$                & $0.51 \pm 0.02$   & $0.55 \pm 0.03$   & $0.73 \pm 0.02$   \\ \midrule
\multirow{4}{*}{Central crystal}
    & Mean \ac{SSIM}             & $0.968 \pm 0.003$ & $0.958 \pm 0.004$ & $0.975 \pm 0.003$ \\
    & $\text{Pct}_{<\tau}$ [\%]  & $0.36 \pm 0.07$   & $0.8 \pm 0.2$     & $0.00 \pm 0.01$   \\
    & $\text{P5}$                & $0.89 \pm 0.02$   & $0.85 \pm 0.03$   & $0.89 \pm 0.02$   \\
    & $\text{P1}$                & $0.36 \pm 0.02$   & $0.23 \pm 0.02$   & $0.72 \pm 0.03$   \\ \bottomrule
\end{tabular}
\label{tab:attribution_ssim}
\end{table}

\begin{figure}[htb]
\centering
\subfloat[Below-threshold \ac{SSIM} regions of the \ac{COG} distribution generated by optiGAN+Fourier.]{%
\includegraphics[width=0.40\textwidth]{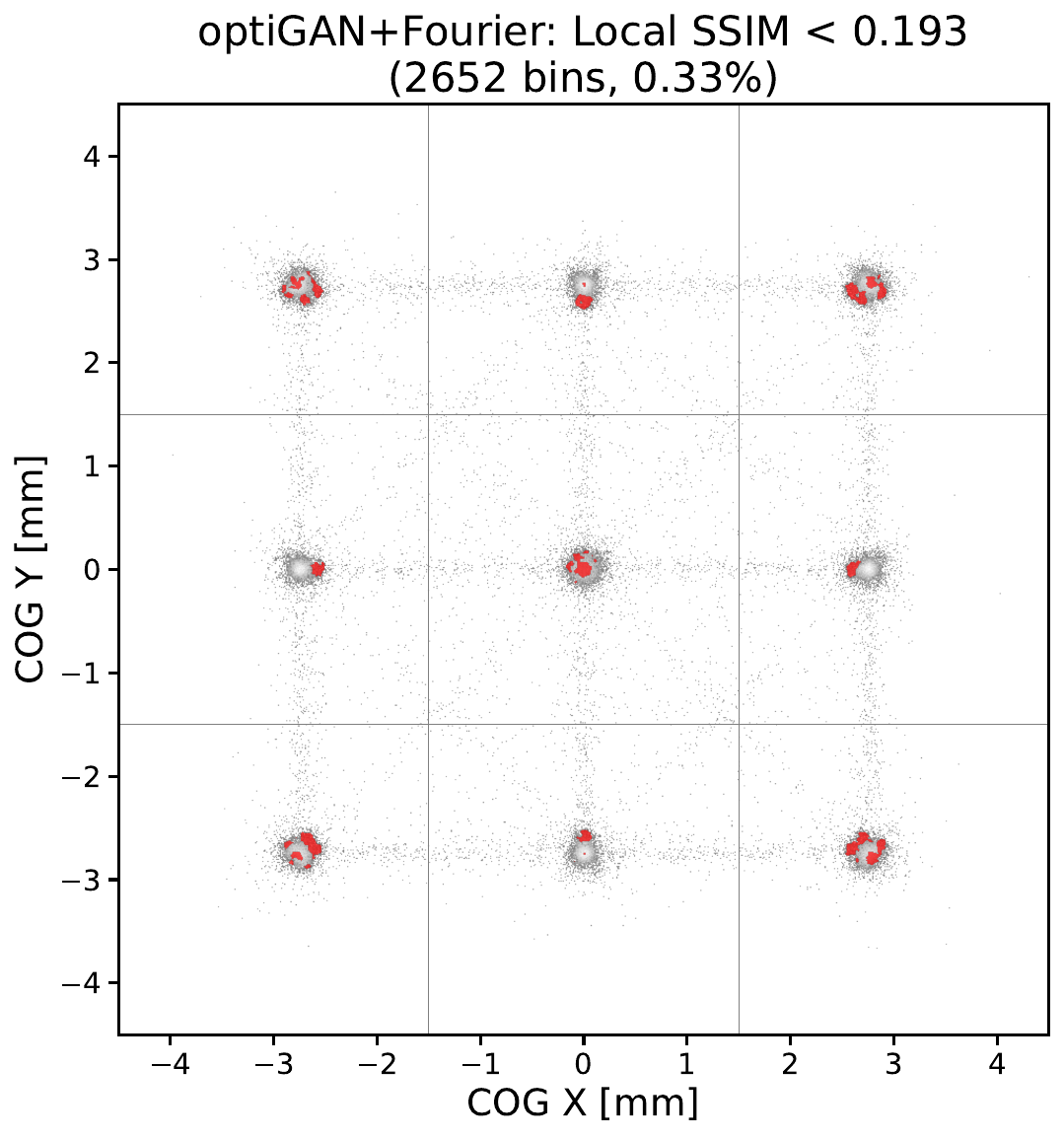}
}
\quad
\subfloat[Below-threshold \ac{SSIM} regions of the \ac{COG} distribution generated by optiGAN+Fourier+PI.]{%
\includegraphics[width=0.40\textwidth]{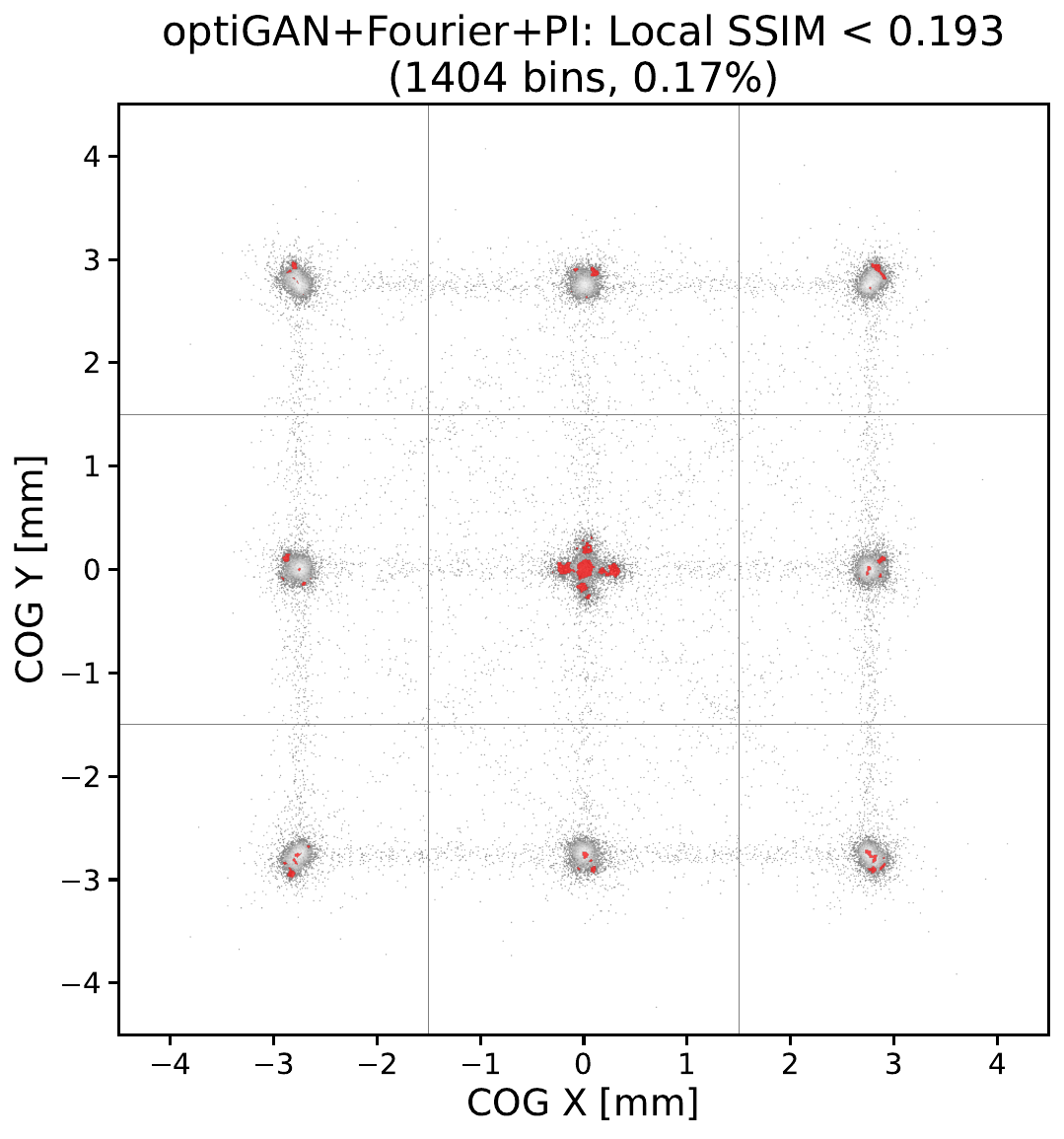}
}
\caption{Spatial distribution of local \ac{SSIM} bins below the threshold $\tau = 0.193$. The left image shows the data generated by optiGAN+Fourier (2652 bins, \qty{0.33}{\percent}). The right image shows the data generated by optiGAN+Fourier+PI (1404 bins, \qty{0.17}{\percent}). Note the redistribution of the affected bins from the eight outer crystals in the Fourier-only case to the central crystal in the full model.}
\label{fig:attribution_lowssim}
\end{figure}

The fraction of bins below the reference threshold is reduced from \qty{0.33 \pm 0.04}{\percent} to \qty{0.17 \pm 0.05}{\percent}, corresponding to a \qty{48}{\percent} reduction of the low-\ac{SSIM} bin count (see also \cref{fig:combined_local_SSIM} in the appendix) when using the unit-norm loss. The regional decomposition reveals that this improvement is not uniformly distributed but concentrated on the eight outer crystals, where the low-\ac{SSIM} bin fraction is reduced by a factor of \num{3.6} (from \qty{0.32 \pm 0.04}{\percent} to \qty{0.09 \pm 0.04}{\percent}), and the mean local \ac{SSIM} increases from \num{0.980 \pm 0.002} to \num{0.982 \pm 0.002}. The per-crystal breakdown (see \cref{tab:attribution_ssim_percrystal} in the appendix) further shows that this outer-crystal improvement is homogeneous across the eight positions. On the central crystal, the same loss term produces the opposite effect. The low-\ac{SSIM} bin fraction increases by a factor of \num{2.2} (from \qty{0.36 \pm 0.07}{\percent} to \qty{0.8 \pm 0.2}{\percent}), and the mean local \ac{SSIM} decreases from \num{0.968 \pm 0.003} to \num{0.958 \pm 0.004}.\newline
\Cref{fig:attribution_lowssim} shows the spatial distribution of bins with local \ac{SSIM} below the reference threshold $\tau$ for both configurations. In the Fourier-only configuration, the affected bins are distributed across all nine photopeak clusters, with the largest counts observed at the four corner crystals. In the optiGAN+Fourier+PI configuration, the number of affected bins on the eight outer crystals drops substantially, while the central-crystal cluster becomes the single dominant contribution.

\section{Discussion}
\label{sec:discussion}
The ablation study (see \cref{subsec:ablation_results}) isolates the network modifications introduced to adapt the single-crystal optiGAN to the \numproduct{3x3} array. Trained on the array data with no modifications, the single-crystal architecture failed to reach a stable training regime, with an increasing fraction of conditions collapsing in condition-wise \ac{SWS} over time. The addition of the Fourier feature encoding together with the latent mapping network restored convergence, establishing them as the technical prerequisite for array-level training.\newline
The additional unit-norm loss term leverages physics by enforcing the unit-sphere state space $S^2$ of propagation directions. It addresses photon transport near crystal-to-crystal boundaries, which is essential in detector arrays. Reflection and transmission at those interfaces produce geometrically complex directional distributions, guidance on momentum phase space factorization helps the model to capture the direction relationships. The region-stratified analysis shows that the momentum-norm error was reduced most strongly for boundaries between adjacent crystals, while transition zones remained unchanged (see \cref{tab:PI-Loss}).\newline
The detection position stratified view (see \cref{fig:MAE-Comp}) shows that this effect is spatially structured. Besides the reduction along the inter-crystal interfaces, greater error is observed in the center of the sensor plane. Because the momentum-norm error is evaluated per generated sample against the analytical ground truth $\|\hat{\mathbf{P}}\|^2 = 1$, this redistribution is directly observable at the sample level. Similar assessment for the position-based flood-map reconstruction is only possible at the distribution level per emission condition and is the subject of the attribution analysis in \cref{tab:attribution_ssim}.\newline
The results of the full array evaluation demonstrate that the model is able to generate optical photon data in good agreement with the ground truth simulation data. Since the similarity performance values between conditions located in the fundamental domain do not differ from conditions outside, the evaluation underlines the efficacy of our training approach. Although one could expect that the generation of high-fidelity data might be the easiest for conditions in the central region, results indicate that the similarity performance is decreased in that region compared to the side and corner crystals. Furthermore, there is a slight indication that the model works better for conditions located in $Z$-layers further away from the photosensor, pronounced through higher \ac{SWS} values. This behavior can be explained by physics, since the spatial components conditioned for farther $Z$-layers show greater uniformity and symmetry.\newline
The high-resolution experiment demonstrates the capabilities of the model to generate high-fidelity optical photon data also for conditions never seen by the network. Results show that points near and far from training conditions achieve a similarity comparable to the ground truth. Similar to the full experiment, it is observed that the lowest similarity values are obtained for the central crystal. The results of the flood map artifact study indicate that this is caused by the unit-norm loss, which improves the generation on the outer eight crystals and crystal boundaries, but degrades performance for central crystal conditions. Regarding different $Z$-layers, the model's performance on conditions located at known ($\qty{1}{\milli \metre}, \qty{5}{\milli \metre}, \qty{9}{\milli \metre}$) and unknown ($\qty{3}{\milli \metre}, \qty{7}{\milli \metre}$) crystal heights shows no significant differences ($<3\sigma$). However, a reduction of the \ac{SWS} values at unknown layers is observed and quantified in \cref{tab:doi_layerwise_sws}, which remains within the intrinsic layer-to-layer variability of the trained-layer ensemble.\newline
The pencil beam evaluation analyzes the performance under common visual characteristics used in experimental radiation detector research. In addition to the final model configuration (optiGAN+Fourier+PI), we also evaluated the Fourier-only version (optiGAN+Fourier). While both flood maps agree within $3\sigma$ with the Monte Carlo base line concerning the mean local \ac{SSIM}, they visually show different features. The flood map of data generated by the optiGAN+Fourier+PI model displays a cross artifact for the center-crystal photocluster. This artifact is absent for the flood map data generated by the optiGAN+Fourier model, although especially the corner photopeak clusters appear bigger in size. Since both models were trained on the identical training dataset comprising only conditions from the fundamental domain, this also indicates that the artifact behavior is not rooted in selecting the fundamental domain as the training region. Visually comparing the lines connecting the cluster spots vertically and horizontally, the line structure seems less defined for the optiGAN+Fourier case.\newline
Although the artifact is visually dominant due to its prominent position in the center, the region-stratified \ac{SSIM} values report overall positive effects of the unit-norm loss. While the loss term produces a localized degradation in the central crystal, where the low-\ac{SSIM} bin fraction increases by a factor of \num{2.2}, a statistically clear improvement in flood-map fidelity on all other crystals is observed, reducing the low-\ac{SSIM} bin fraction by a factor of \num{3.6}. The net effect over the full array is a \qty{48}{\percent} reduction of the low-\ac{SSIM} bin count.\newline
The localization of this effect at the central crystal is not fully explained by the present data. Two possible causes can be excluded. First, although the $D_4$ training grid with a fixed step width of \qty{0.6}{\milli \metre} has fewer conditions within the central crystal, the optiGAN+Fourier baseline trained on this grid does not produce the artifact, which indicates that training density and asymmetry are not the cause. Second, the central crystal is the only crystal fully surrounded by neighbors, which could produce a different physical distribution of photon momenta than the outer crystals. However preliminary analyses of the marginal and outward-momentum components in different crystal regions show no such differences and thus rule out a boundary-related selection effect. An explanation for the localization at the central crystal, and why it doesn't appear in the Fourier-only model, remains an open question for future work.

\section{Summary \& Outlook}
In this work, we presented an array-level extension of optiGAN and demonstrated its ability to generate optical photon data within a \numproduct{3 x 3} \ac{BGO} crystal array, using our recent LUT Davis model of optical crosstalk in pixelated detectors \cite{trigila_intercrystal_2024}. We introduced Fourier feature encoding together with a latent mapping network as the technical prerequisite for stable training on the array geometry, and a unit-norm loss term enforcing the $S^2$ state space of propagation directions as a soft constraint, which constitutes the physics-informed contribution of this work. Both modifications preserve the generator and discriminator architectures of the single-crystal optiGAN, keeping the computational cost tractable and the model accessible to the GATE user community.\newline
The model produced optical photon data in good agreement with Monte Carlo ground-truth simulations, with strong generalization and interpolation capabilities across emission positions and depth-of-interaction layers not present during training. The unit-norm loss term improves the flood-map fidelity on all outer crystals and crystal boundaries (reduction of low-\ac{SSIM} bins by a factor of \num{3.6}), but is also accompanied by a localized degradation at the central crystal (increase by a factor of \num{2.2}), yielding a net \qty{48}{\percent} reduction of low-\ac{SSIM} bins over the full array. The origin of the localization at the central crystal specifically, and why it does not occur in the Fourier-only model, is not fully explained by the present data. Training-density asymmetry and differential physical distributions were excluded on the basis of the ablation study and preliminary momentum-distribution analyses, and a definitive explanation remains an open question.\newline
Future work will proceed in different directions. Hyperparameter optimization and deeper network architectures are expected to improve fidelity for users with access to more powerful hardware than the present setup, focused on broad accessibility. Larger arrays such as \numproduct{5 x 5} or \numproduct{11 x 11}, in which the fully-surrounded and $D_4$-invariant crystal positions no longer coincide, would both allow disentangling the mechanism behind the central-crystal localization and extend the model toward practical array sizes. Furthermore, we aim at finding a universal recommendation for the training grid design that suits different array configurations and crystal dimensions. For the future, we will explore conditioned generation of specific physics scenarios, such as constraining inter-crystal scattering to selected regions, which is difficult to achieve in classical Monte Carlo simulations.

\section{Acknowledgment}
This work is supported by NIBIB grant R01 EB034475.

\newpage
\section*{Appendix}
\begin{table}[H]
\caption{Per-crystal region-stratified local \ac{SSIM} statistics for optiGAN+Fourier and optiGAN+Fourier+PI. Values are point estimates $\pm$ bootstrap standard deviation ($N_\text{bootstrap} = 100$), rounded such that the uncertainty carries one significant figure. Crystal labels follow a 3$\times$3 grid convention: B/M/T denote the bottom, middle, and top row; L/C/R denote the left, center, and right column. $\text{Pct}_{<\tau}$ denotes the fraction of bins in the local \ac{SSIM} map falling below the threshold $\tau = 0.193$. $\text{P5}$ is the 5th percentile of the local \ac{SSIM} distribution.}
\centering
\begin{tabular}{@{}llcc@{}}
\toprule
Configuration & Crystal & $\text{Pct}_{<\tau}$ [\%] & $\text{P5}$ \\ \midrule
\multirow{9}{*}{optiGAN+Fourier}
    & BL & $0.47 \pm 0.06$ & $0.93 \pm 0.01$ \\
    & BC & $0.13 \pm 0.03$ & $0.91 \pm 0.02$ \\
    & BR & $0.53 \pm 0.06$ & $0.93 \pm 0.01$ \\
    & ML & $0.15 \pm 0.04$ & $0.91 \pm 0.02$ \\
    & MC & $0.36 \pm 0.07$ & $0.89 \pm 0.02$ \\
    & MR & $0.15 \pm 0.04$ & $0.91 \pm 0.02$ \\
    & TL & $0.44 \pm 0.08$ & $0.93 \pm 0.01$ \\
    & TC & $0.20 \pm 0.03$ & $0.91 \pm 0.02$ \\
    & TR & $0.52 \pm 0.07$ & $0.93 \pm 0.01$ \\ \midrule
\multirow{9}{*}{optiGAN+Fourier+PI}
    & BL & $0.11 \pm 0.05$ & $0.93 \pm 0.01$ \\
    & BC & $0.08 \pm 0.06$ & $0.91 \pm 0.02$ \\
    & BR & $0.14 \pm 0.04$ & $0.93 \pm 0.01$ \\
    & ML & $0.08 \pm 0.06$ & $0.91 \pm 0.02$ \\
    & MC & $0.8 \pm 0.2$   & $0.85 \pm 0.03$ \\
    & MR & $0.09 \pm 0.05$ & $0.91 \pm 0.02$ \\
    & TL & $0.06 \pm 0.05$ & $0.94 \pm 0.01$ \\
    & TC & $0.08 \pm 0.05$ & $0.91 \pm 0.02$ \\
    & TR & $0.10 \pm 0.05$ & $0.94 \pm 0.01$ \\ \bottomrule
\end{tabular}
\label{tab:attribution_ssim_percrystal}
\end{table}

\begin{figure}[htb]
\centering
\subfloat[Distribution of the local \ac{SSIM} values.]{%
\includegraphics[width=0.45\textwidth]{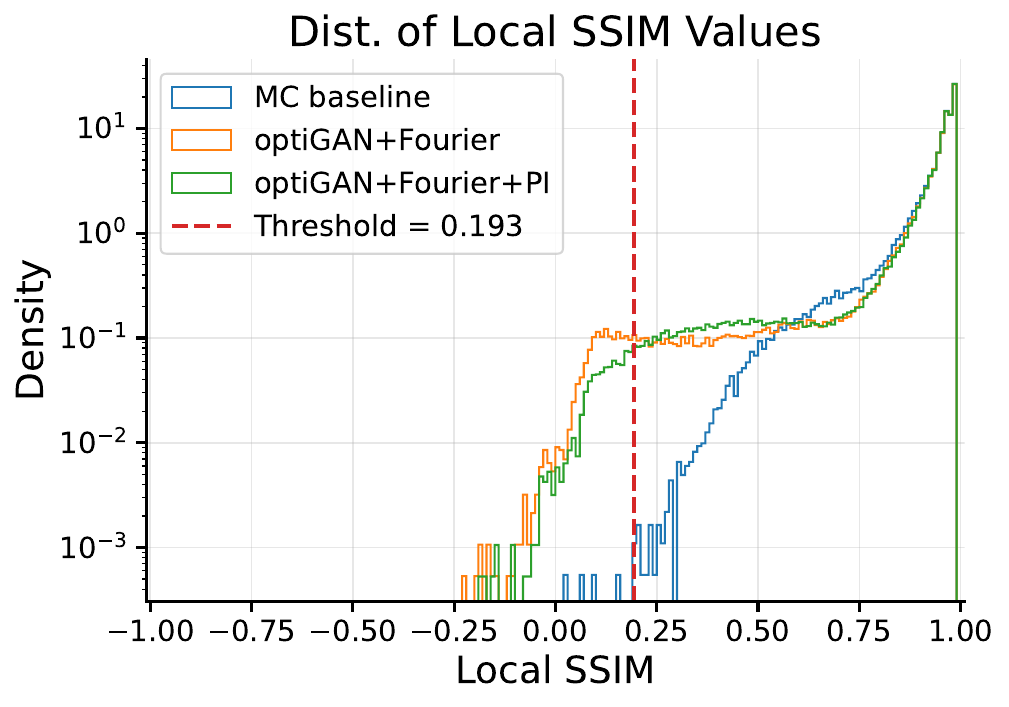}
}
\quad
\subfloat[Cumulative distribution of the local \ac{SSIM} values.]{%
\includegraphics[width=0.45\textwidth]{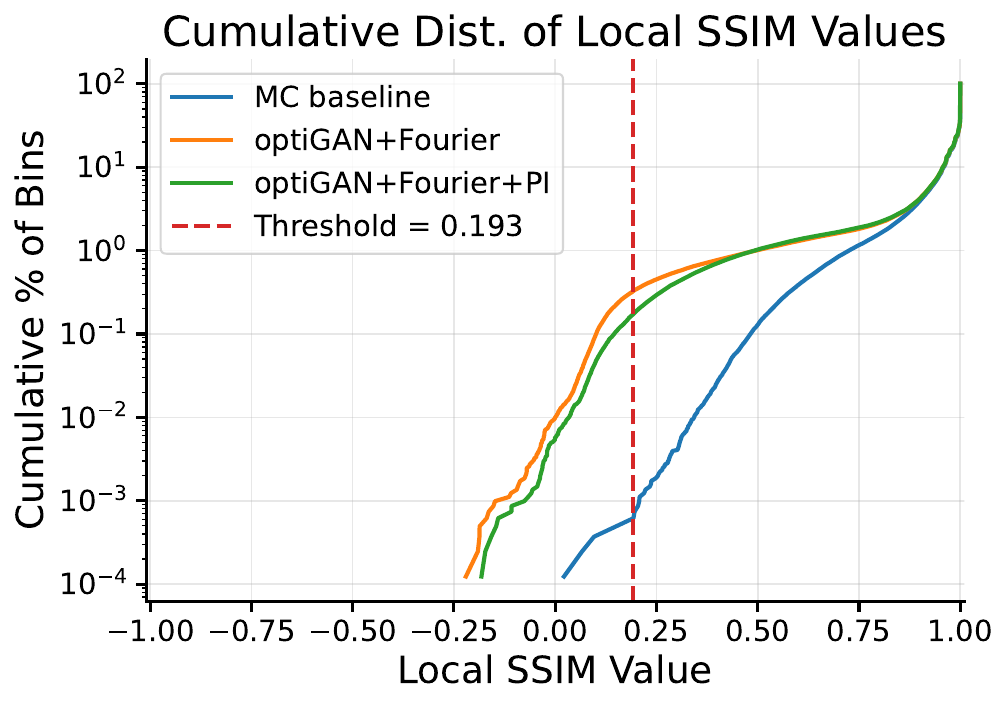}
}
\caption{Local \ac{SSIM} distribution of optiGAN+Fourier, optiGAN+Fourier+PI and the Monte Carlo baseline. The set of local \ac{SSIM} values is given by the \ac{SSIM} estimate computed within a sliding window centered at each pixel location, capturing spatially resolved structural similarity across the image. The \ac{SSIM} threshold was chosen to be equal to the lower end of the baseline simulation's distribution tail.}
\label{fig:combined_local_SSIM}
\end{figure}

\newpage
\bibsetup
\printbibliography

\newpage

\end{document}